\documentclass[journal]{IEEEtran}
\ifCLASSINFOpdf
\else
\fi

\usepackage{cite}
\usepackage{amsmath}
\usepackage{amsfonts}
\usepackage{graphicx}
\usepackage{ amssymb }
\usepackage{xcolor}
\usepackage{bbold}
\usepackage{cite}
\usepackage{subcaption}		
\usepackage{arydshln}
\usepackage{enumitem}

\newcommand{\R}{\mathbb{R}}

\newcommand{\C}{\mathbb{C}}
\newcommand{\Z}{\mathbb{Z}}
\newcommand{\F}{\mathcal{F}}

\newcommand{\h}{\mathcal{H}}

\newcommand{\first}{1^{\text{st}}}

\newcommand{\bbm}{\begin{bmatrix}} 
\newcommand{\ebm}{\end{bmatrix}} 
\newcommand{\lb}{\left(} 
\newcommand{\rb}{\right)} 
\newcommand{\mbo}{\mathbb{1}}
\newcommand{\req}[1]{(\ref{#1.eq})}

\newcommand{\cK}{{\cal K}}

\newcommand{\sm}{\text{-}}
\newcommand{\ssT}{{\scriptscriptstyle T}}
\newcommand{\ssN}{{\scriptscriptstyle N}}
\newcommand{\ssM}{{\scriptscriptstyle M}}
 
\newcommand{\cA}{{\cal A}}

\newcommand{\cS}{{\cal S}} 
\newcommand{\cL}{{\cal L}} 

\newcommand{\lcb}{\left\{} 
\newcommand{\rcb}{\right\}}
\newcommand{\fdbk}[2]{{\cal F} \! \left( #1 ; #2 \right)}

\newcommand{\Ptilde}{\tilde{P}}
 
\newcommand{\spM}[1]{\mathfrak{sp}\!\left( #1 \right)} 
\newcommand{\sA}{\mathcal{S}_{\cal A}} 
\newcommand{\skw}{\mathcal{S}} 
\newcommand{\Ims}[1]{{\rm Im}\!\left( #1 \right) } 
\newcommand{\Nus}[1]{{\rm Nu}\!\left( #1 \right) } 
\newcommand{\inprod}[2]{\left< #1 , #2 \right>} 
\newcommand{\diag}[1]{ {\sf diag} \! \left( #1 \right) \rule{0em}{1em}}

\newcommand{\one}{\mathbb{1}}

\newcommand{\ocA}{{\cA}}

\newcommand{\Phix}{\Phi_{xx}}
\newcommand{\Phiu}{\Phi_{ux}}

\newcommand{\lba}{\left[ \begin{array}}
\newcommand{\ear}{\end{array} \right]}
\newcommand{\be}{\begin{equation}} 
\newcommand{\ee}{\end{equation}} 
\newcommand{\ba}{\begin{aligned}} 
\newcommand{\ea}{\end{aligned}}

\newtheorem{thm}{Theorem}[section]
\newtheorem{lem}[thm]{Lemma}
\newtheorem{prop}[thm]{Proposition}
\newtheorem{cor}[thm]{Corollary}
\newtheorem{defn}{Definition}[section]

\newtheorem{exmp}{Example}[section]
\newtheorem{rem}{Remark}

\begin{document}

%
\title{On Structured-Closed-Loop versus Structured-Controller Design: \\ the Case of Relative Measurement Feedback
}
%
%
%

\author{Emily Jensen$^{1}$ and Bassam Bamieh$^{2}$
\thanks{This work is partially supported by NSF award ECCS-1932777 and CMI-1763064.}
\thanks{$^{1}$Emily Jensen is with the department of Electrical and Computer Engineering,
      University of California, Santa Barbara
        {\tt\small emilyjensen@ucsb.edu}}%
\thanks{$^{2}$Bassam Bamieh is with the Department of Mechanical Engineering, University of California, Santa Barbara
        {\tt\small bamieh@ucsb.edu}}%
}

\maketitle

\begin{abstract}
We consider the optimal distributed controller design problem subject to two structural requirements: {\em locality}, i.e. available measurements and sub-controllers' interactions are governed by a graph structure, and {\em relative feedback}, i.e. only differences of measurements are available to the controller. 
We formalize controller locality in terms of the controller's transfer function, state-space realization, or resulting closed-loop mapping. We demonstrate that the relative feedback requirement can be written as a convex constraint on the controller and (in special cases) on the resulting closed-loop, and we characterize the allowable structures of relative feedback controllers. 
We prove that sparse closed-loop design is a convex relaxation of structured controller state-space design, even in the continuous time IIR setting. This formalizes and extends results of the recently developed System Level Synthesis framework. We take a first step toward quantifying the performance gap associated with this convex relaxation by constructing a class of examples (based on relative feedback requirements) for which the difference in performance, measured by an $\h_2$ norm, is infinite.
 The results presented are used to contrast several issues of structural constraints in distributed control design that remain as open problems. 

 \end{abstract}


%
\IEEEpeerreviewmaketitle

\section{Introduction}

It is generally believed that optimal and robust control design for LTI systems is a mature area of research. Such problems include  Linear Quadratic Gaussian (LQG or $\h^2$), $\h^\infty$ and $\ell^1$ optimal designs. However, in many applications additional structural constraints on these problems must be imposed. With constraints imposed, optimal controller design generally becomes significantly more challenging, and satisfactory solutions to many structurally constrained problems remain elusive.
Examples include decentralized control,  fixed-order, static output-feedback, and stable/stabilizing optimal controller design. Due to the importance of these design criteria, various proxy solutions have been devised, e.g. controller reduction schemes have been proposed as a proxy for fixed-order optimal controller design.

An increased prevalence of distributed control over the past two decades has renewed interest in constraints on the {\em spatial structure} of the controller. These constraints correspond to the fact that in a networked setting, spatially-distributed sub-controllers typically access only a \emph{local} subset of  system information and communicate to a restricted neighboring subset of  components. Even if access to full information is possible, it may be intractable to implement a fully centralized policy for very large scale systems. Currently, there are a plethora of such design constraints that are termed localized, networked structured, etc.

 Despite a more lengthy history of the emphasis of locality structure in distributed control literature, 
 specific attention on how to {\em realize} (i.e. implement) distributed controllers in a local way has only quite recently been regarded as a key issue~\cite{vamsi2015optimal,vamsi2011optimal,lessard2013structured,rantzer,naghnaeian2018unified}. 
For example, a controller with tridiagonal $A$, $B$, $C$, and $D$ matrices comprising its state space realization can be viewed as 
a natural realization for a distributed controller composed of several sub-controllers arranged on a one-dimensional lattice with access only to neighboring sub-controllers' states and plant measurements. The corresponding controller transfer function matrix will generally not be tridiagonal but rather a full transfer matrix (though with some other structure on relative degree of entries). On the other hand, a tridiagonal transfer function matrix can indeed always be realized by tridiagonal state space matrices. Thus the set of controllers with structured transfer matrices does not fully capture the set of controllers with realization matrices of the same sparsity structure. Moreover, recent work on this topic has uncovered that minimal realizations are not necessarily desirable in the distributed setting, where the complexity of the controller is not simply expressed by the overall state dimension. Instead there appears to be a tradeoff between state dimension and the ability to realize controllers in a pre-specified distributed manner. This ``spatially structured realization problem'' remains open, and the results of this paper motivate its further study.

In addition to locality, another important structural constraint on controllers is one we refer to as {\em relative feedback}. Such constraints are implicit in many multi-agent systems problems with sensors that measure only relative quantities, or more precisely, differences between outputs of the plant, but to our knowledge have not been studied as an explicit design constraint. Examples include phase differences to determine the interaction between coupled oscillators in 
AC power systems, relative strain measurements for mechanical applications, relative distance measurements between vehicles in 
vehicular formations  \cite{jovanovic2005ill}, satellite constellations \cite{psiaki1999autonomous}, and between mirror segments in 
the European Extremely Large Telescope~\cite{sarlette2014control}.
Imposing relative feedback during design is also beneficial as relative sensing devices tend to be simpler to implement  than sensors that provide absolute measurements.

It is known that restriction to relative sensors (as opposed to ones that measure the actual outputs) may impose fundamental limitations on the performance of large-scale systems, typically on the ability to control spatially large-scale (i.e. nonlocal) modes of low temporal frequencies.
For example, vehicular platoons with only relative measurement devices have been shown~\cite{bamieh2012coherence} to have arbitrarily degrading best-achievable-performance as the system size increases. This motivates the in depth study of relative feedback as a design constraint.

Local controller design and relative feedback design are specific examples of structurally constrained controller design problems.
Solutions to structured controller design have been developed for a few special problem settings. 
One such setting is when the problem is convex. Examples include partially nested information structures~\cite{1099850} and generalizations~\cite{voulgaris1999control}, funnel causality~\cite{bamieh2005convex} and quadratic invariance~\cite{rotkowitz2005characterization}. An intuitive result~\cite{bamieh2005convex,rotkowitz2005simple} in this setting is that if the structural constraints allow information within the controller to travel ``faster'' than disturbances in the plant, then the constrained optimal control problem is convex. 

 In another special setting, the structure of the (unconstrained) 
 optimal controller (state dimension, time invariance, spatial symmetries, approximate locality) comes ``for free'', inherited from the plant and performance objective. 
For example, optimal $\h^2$ and $\h^\infty$ controllers have at most the same state dimension as the plant, and time-varying controllers do not outperform time-invariant disturbance attenuation controllers for LTI plants~\cite{shamma1989time,feintuch1985uniformly}. 
In the distributed setting, optimal controllers inherit the spatial symmetries of the  system and  performance criterion~\cite{fagnani1993representations,bamieh2002distributed}.
For idealized spatially-invariant plants with  infinite spatial extent and locally interacting subcomponents, optimal controllers are ``almost localized'' in the sense that controller gains decay exponentially in space~\cite{bamieh2002distributed,motee2008optimal,motee2017sparsity}. 

Many problems do not fall into either of the two special categories mentioned above, motivating further study of methods for structured controller design. For instance, 
although controller gains for spatially-invariant problems decay exponentially in space, the decay rates are determined by the plant's dynamics and the performance metric \cite{bamieh2002distributed} with no known tractable procedure to obtain faster decay. 
The literature on structured controller design is rapidly growing, so we only mention that which is directly related to the present work. The most recent of those is the System Level Synthesis (SLS) framework~\cite{wang2019system,matni2017scalable,anderson2019system}, which has recently gained a large amount of attention in the literature in both theory \cite{wang2014localized , matni2017scalable} and applications \cite{alonso2019distributed, li2020mpc, dean2020robust}.

The SLS framework is a combination of two principles. The first follows from the fact that for linear systems, the set of all stabilized closed loops is affine-linear. Thus any convex structural constraint on the {closed loop} preserves convexity of the design problem. For example, a fixed band size constraint on the closed loop transfer function matrix (or more general pre-specified sparsity structure constraints) are convex. Thus, we also refer to SLS in this paper as ``closed-loop design''.

 Imposing a structural constraint on the closed loop is of course not equivalent to imposing the same constraint on the controller (except in the cases of funnel causality or quadratic invariance). 
Thus the second principle of SLS is to demonstrate that the controller can be ``implemented" using components of the designed closed-loop, preserving the constrained structure. 
In this way the SLS procedure can be thought of as a convex relaxation of the structured controller ``implementation" design problem. 
The SLS procedure is both theoretically elegant and has compelling computational complexity advantages~\cite{wang2019system,matni2017scalable} which render controller computations for large-scale systems tractable. Therefore it is important to formalize this idea of structured controller ``implementation" and extend this idea to continuous time IIR settings, as well as quantify the potential performance gap between SLS and the structured controller design problem. These are some of the problems addressed in this paper.


\subsection*{Contributions \& Paper Structure}

The paper is structured as follows: 
In Section~\ref{sec:setup} we introduce the structured controller design problem and motivate two types of structural constraints, {\em locality} and {\em relative feedback}, through an example. We formalize three possible notions of locality in Section~\ref{sec:locality}. In Section~\ref{sec:convexRelaxation}, we demonstrate that {\em sparse closed-loop design is a convex relaxation of the sparse controller realization design problem.} We next provide a framework for relative feedback as a design constraint in Section~\ref{sec:relative_feedback}. In particular, we provide a {\em characterization of  the structures of relative feedback controllers} and demonstrate that {\em relative feedback can be posed as a convex constraint on the closed-loop} in certain problem settings. In Section~\ref{sec:gap} we construct a class of examples based on relative feedback constraints for which {\em the performance gap between the sparse controller realization design problem and the convex relaxation provided by sparse closed-loop design is infinite}. We end with a discussion of a few of the many remaining open questions in this area.

Preliminary results were presented in \cite{jensen2020Gap}. These results did not include analysis of closed-loop structure in the output feedback setting, the characterization of relative feedback controller structure, or analysis on the effect of graph connectivity. The performance gap presented in preliminary work relied on spatial-invariance of the controller and applied only to one specific performance measure.

\section{Problem Set-up \& Motivation}   \label{sec:setup}
In this section, we set up the structurally constrained controller design problem and provide a motivating example. 

		We consider a generalized plant $P$ partitioned as 
			\be 
		\bbm z \\ y \ebm = \bbm P_{11} & P_{12} \\ P_{21} & P_{22} \ebm 
		\bbm w \\ u \ebm ,
	 \label{Gen_Plant.eq}
	\ee
	where the vector-valued signals $w$, $u$, $y$ and $z$ are the exogenous disturbances, control signals, 
	 measurements available to the controller, and performance output respectively. 
	 $P$ corresponds to a spatially-distributed system, which means that 
	 the signal vectors $w$, $u$,  $y$ and $z$ are partitioned into local sub-signals, e.g.
	\be
		u ~=~ \bbm u_1^\ssT & \cdots & u_\ssN^\ssT \ebm^\ssT , 
	  \label{u_partition.eq}
	\ee
	where $u_n$ is the control signal at the  $n^{\rm th}$ site. State space realizations of $P$ are of the form
		\be \ba \label{eq:plantSS}
			\dot{x} &=  A x+ B_1 w + B_2 u \\
			z & = C_1 x + D_{12} u\\
			y & = C_2 x + D_{21} w, 
		\ea \ee
	and are denoted compactly as 
	 \begin{equation}  \label{eq:linearDynamics}
		P =  \left[ \begin{array}{c|cc} A & B_1 & B_2 \\ \hline C_1 & 0 & D_{12} \\ C_2 & D_{21} & 0 \end{array} \right] 
		=  \left[ \begin{array}{cc} P_{11} & P_{12} \\ P_{21} & P_{22} \end{array} \right].	
	\end{equation} 	
	The state $x$ is partitioned similarly to~\req{u_partition}, and partitioning of signals induces partitionings of the
	 realization matrices $A, B_i, C_i, D_{ij}$.
	 
\begin{figure}
\setlength\belowcaptionskip{-.8\baselineskip}
	\begin{centering}
	\includegraphics[width=0.25\textwidth]{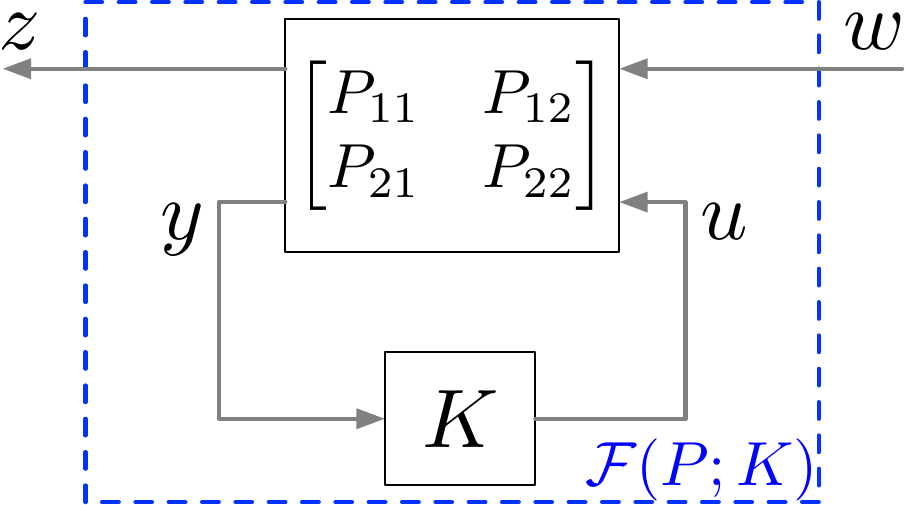}
 	 \caption{\footnotesize Feedback interconnection of plant $P$ with LTI output feedback controller $u = Ky$. The closed-loop mapping from exogenous disturbance $w$ to performance output $z$ is denoted by $\mathcal{F}(P;K)$.
	  }
	 \label{fig:fdbk}
	 \end{centering}
\end{figure}

		When plant $P$ is in feedback with a (possibly dynamic) LTI  output feedback controller
			$$
				u = Ky, 
			$$
		the closed-loop mapping from exogenous disturbance $w$ to performance output $z$ is given by 
			\be \label{eq:cl}
		\fdbk{P}{K} := P_{11} + P_{12} K \lb I - P_{22} K \rb^{-1} P_{21}, 
			\ee
		(see Figure~\ref{fig:fdbk}). 
		The design problem of interest is to find $K$ which minimizes a closed-loop norm.
In distributed settings, $K$ must satisfy additional structural requirements over classical controller synthesis to conform with the limited sensing and communication architecture:
	\begin{subequations} \label{eq:cl_opt_const}
	\begin{align}
		&\inf_K ~\| \fdbk{P}{K} \| \label{eq:design_cost}\\
		&~ {\rm s.t.} ~\text{structural requirements} \label{eq:struct_require}
	\end{align}
	\end{subequations}
When there does not exist any $K$ satisfying the constraints \eqref{eq:struct_require} that results in a finite cost \eqref{eq:design_cost}, we say that problem \eqref{eq:cl_opt_const} is {\em infeasible}.

\subsection{Motivating Example: Consensus of $1^{\rm st}$-Order Subsystems} \label{sec:consensus} 
In this section, we present an example that motivates the concepts studied throughout this paper. 
Consider the $\h_2$ controller design problem for a plant composed of $N$ decoupled $\first$-order subsystems:
	\be \label{eq:plantState}
		 \dot{x}_n(t) = u_n(t) + w_n(t),~ n \in \Z_N,
	\ee
where $x_n, u_n, w_n$ are the state, control action, and local disturbance at spatial location $n$, respectively.
$x$, $u$ and $w$ denote the vectorized state, control and disturbance e.g. $x = \bbm x_0 & \cdots & x_{\ssN-1} \ebm^\ssT$. The performance output is 
	\begin{equation}  \begin{aligned} \label{eq:PlantDynamics}
		z&= \left[ \begin{array}{c} C \\ 0 \end{array} \right] x + \left[ \begin{array}{c} 0 \\ \gamma I \end{array} \right] u,
	\end{aligned} \end{equation}
where $C$ is chosen so that the $2$-norm of the signal $Cx$ captures a measure of consensus. We consider $C$ corresponding to  a ``deviation from average'' metric:

	\be \label{eq:ave_metric}
		(C^{\rm ave.}x)_n = x_n - \frac{1}{N} \sum_{i=1}^N x_i.
	\ee
 The closed-loop $\fdbk{P}{K}$ provides a consensus algorithm. A standard solution is given by:
\be  
	u_n ~=~  k_{+} \big(  x_{n+1}-x_n \big) ~+~  k_{-} \big(  x_{n-1}-x_n \big) , 
  \label{eq:standard_consensus}	
\ee
	which is easily shown to result in finite closed-loop norm $\| \fdbk{P}{K} \|_{\h_2}$ when $k_{+}, k_{-}> 0$. Note that the policy \eqref{eq:standard_consensus} has the following two structural properties:
	\begin{itemize}
		\item {\em Locality:} computation of the control at spatial site $n$ requires access only to information from ``nearby" subsystems ($x_{n-1}, x_n, x_{n+1}$),  
		\item {\em Relative Feedback:} each control action is computed using only differences of system information e.g. $x_n - x_{n-1}$. 
	\end{itemize}

Obtaining an \emph{optimal} controller which satisfies these properties corresponds to solving a problem of the form:
	\begin{subequations} 
	\begin{align} 
		&\inf_K ~\| \fdbk{P}{K} \|_{\h_2}^2\\
		&~ {\rm s.t.} ~\text{relative feedback \& locality}
	\end{align}\label{eq:relative_local}\end{subequations}
When the constraints are removed, the optimizer of \eqref{eq:relative_local} is a static state feedback matrix 
(the LQR solution). In contrast, the constrained problem is 
generally non-convex and likely to have an optimizer that is LTI and dynamic. 

We emphasize that there are many ways to formalize the locality constraint here. 
For example, \eqref{eq:relative_local} was analyzed for this consensus example with locality imposed as a banded structure on the controller transfer matrix, under the additional constraint that the controller is static~\cite{bamieh2012coherence} or locally first-order \cite{tegling2019fundamental}.
Best-achievable-performance bounds were obtained as the system size $N\rightarrow\infty$ and were shown to scale  unboundedly with $N$ when the ``band size'' of $K$ remains constant. An open problem is whether such fundamental limitations can be overcome by $K$ with larger state dimension. Motivated by this, we consider whether there is a ``better" way to specify the locality constraint to efficiently search over a larger set of controllers. To this end, in the next section we present three different notions of locality.

\section{Locality Constraints} \label{sec:locality}
In this section, we formalize three notions of locality: {\em (A)} transfer function locality,  {\em (B)} state-space realization locality, and {\em (C)} closed-loop locality. Although each of these notions has separately been considered before, to our knowledge a systematic comparison of these notions has yet to be reported. 
We define each notion with respect to a known underlying communication graph, denoted by $\mathcal{A}$. With some abuse of notation we also let $\cA$ denote the corresponding adjacency matrix.  $\cA^{(b)}$ denotes the
corresponding b-hops graph, i.e. $\cA^{(b)}_{ij} =1$ if there is a path of length $b$ from node $i$ to $j$.

\subsection{Transfer Function Locality} \label{subsec:tf_struct}
Locality constraints have often been specified in terms of the system's input-output mapping (transfer function). This notion of locality is formalized as follows.
\begin{defn} \label{def:A-structured}
An LTI system with $n\times n$ block-partitioned transfer matrix $H$ is {\em transfer-function structured}
		 (\emph{TF-structured}) with respect to graph $\cA$ if 
		\[
			\cA_{ij}=0 ~\Rightarrow~H_{ij}(s) = 0,  
		\]
		where $H_{ij}$ is the $ij$'th block of the transfer function matrix. In other words, $H$ has the same block sparsity pattern as $\cA$.
\end{defn}

\subsection{State-Space Realization Locality}
More recent work has alternatively considered imposing locality constraints in terms of realization matrices \cite{vamsi2015optimal,vamsi2011optimal,lessard2013structured,rantzer,naghnaeian2018unified}.	 \begin{defn} 
         	A matrix $M$ partitioned into $n\times n$ blocks is {\em $\cA$-structured}, denoted by $M\in\cS(\cA)$, if the $ij$'th block of $M$ is zero 
		whenever $\cA_{ij}=0$, i.e. the non-zero blocks in $M$ correspond to edges in the graph.         
	 \end{defn} 
	\begin{defn} \label{def:structuredrealizable}
		A state space realization 
		\be \label{eq:realization}
			H= \left[ \begin{array}{c|c} A & B \\ \hline C & D \end{array} \right], 
		\ee
		of an LTI system $H$
		is an $\cA$-\emph{structured realization} if 
		\be \label{eq:ABCD}
			A,B,C,D \in \cS(\cA) .
		\ee
	When additionally $C$ and $D$ are block diagonal, \eqref{eq:realization} is an $\cA$-\emph{network realization}. If there exists an $\cA$-structured (resp. $\cA$-network) realization of a system $H$, we say that $H$ is $\cA$-\emph{structured-realizable} (resp. $\cA$-{\em network-realizable}). 		\end{defn}

We emphasize that a structured or network realization is likely \emph{not} minimal\footnote{The role of minimality in structured controller realizations will be discussed in Section~\ref{sec:discussion}.}. 
That said, any realization utilized in practice should be stabilizable and detectable, motivating the following terminology. 
\begin{defn}
	A state-space realization of an LTI system is \emph{admissible} if it is both stabilizable and detectable. 
\end{defn}

We omit the explicit graph reference and simply refer to the above notions as TF-structured and structured- and network-realizable when the underlying graph is clear from context.

\subsection{Closed-Loop Locality}						\label{SLS.sec}
A third notion of controller locality is in terms of the resulting closed-loop mappings; this is considered in the recently developed System Level Synthesis (SLS) methodology \cite{wang2019system}, which we sometimes refer to in this paper as closed-loop design. To formalize this notion, we first briefly review one key idea of SLS; for a more detailed presentation see e.g. \cite{anderson2019system}.

\begin{figure}[t]
\setlength\belowcaptionskip{-1\baselineskip}
	\centering 
		\begin{subfigure}[t]{0.45\textwidth}
		\setlength\belowcaptionskip{.5\baselineskip}
		\centering
			\includegraphics[width=0.9\textwidth]{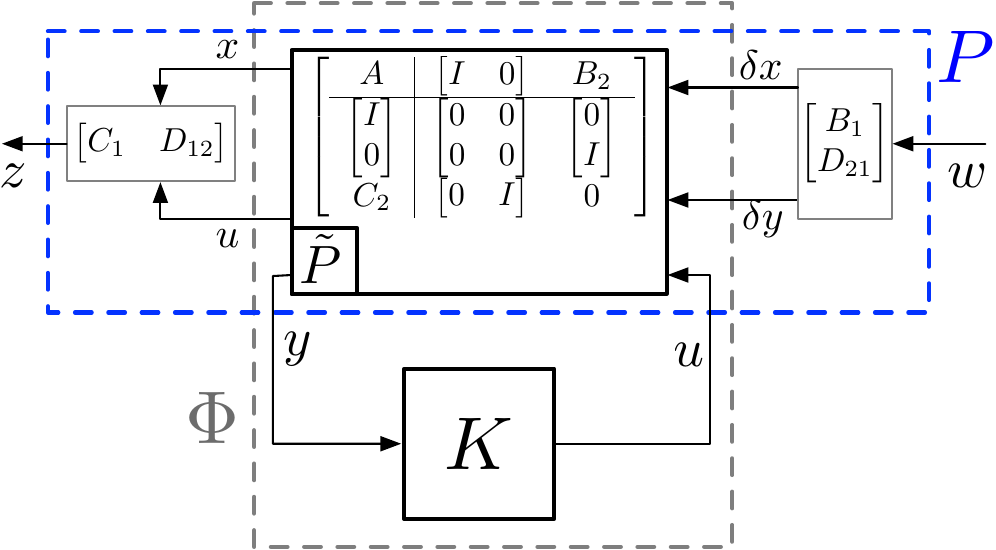}

		\subcaption{\footnotesize
			 Instead of characterizing the affine linear set of all closed loops $\fdbk{P}{K}$ with complicated 
			interpolation constraints, one can 
			characterize the closed loops $\Phi={\cal F}(\Ptilde;K)$, for which the affine linear constraints~\req{CL_Constraints}
			are much simpler. The design problem is then in terms of $\Phi$, but the objective function remains the original one.  } 
	\label{SLS_P_tilde.fig}
	\end{subfigure} 
	
	\bigskip
		
	\centering 
		\begin{subfigure}[t]{0.4\textwidth}
		\setlength\belowcaptionskip{.5\baselineskip}
		\centering
			
			\includegraphics[width=0.7\textwidth]{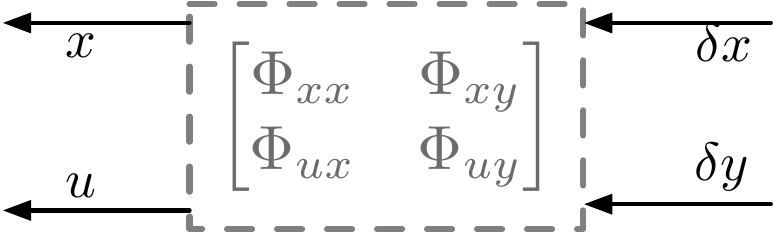}

		\subcaption{\footnotesize The closed-loop $\Phi = \mathcal{F}(\tilde{P},K)$ is partitioned to represent the mappings from disturbances $\delta_x: = B_1 w$ and $\delta_y:= D_{21} w$ to state $x$ and control $u$. }
	 \label{phi.fig}		
	\end{subfigure} 
		
	\caption{\footnotesize The first key idea in the SLS framework is reformulate the controller design problem for $P$ as one for $\tilde{P}$ by characterizing the set of achievable closed-loop maps $\Phi$ rather than the set of achievable $\fdbk{P}{K}$.}
	\label{SLS_Generalized_P.fig}
	
\end{figure}

Consider the plant \eqref{eq:plantSS}
 in feedback with a (possibly dynamic) LTI controller $u = Ky$.
The set of achievable closed-loop mappings $\mathcal{F}(P;K)$ for $P$ is affine linear but is complex to characterize, involving interpolation constraints on the plant's MIMO zeros and their directions \cite{dahleh1994control, boyd1991linear}. 
{\em The first key idea of SLS is to rewrite the controller design problem in terms of }$ \Phi = \mathcal{F}(\tilde{P};K)$ (see Fig.~\ref{SLS_Generalized_P.fig}), leveraging the fact that
characterizing the set of achievable $\mathcal{F}(\tilde{P};K)$ is much simpler:
   \begin{itemize} 
	\item 
	The original controller design problem for $P$ is reformulated as one for $\tilde{P}$ with an equivalent objective:
		$$
		\inf_{K }~	\left\|  \fdbk{P}{K}  \right\| 
		~=~ 	
		\inf_{K } ~\left\| \bbm C_1 & D_{12} \ebm  \mathcal{F}\big(\tilde{P};K\big)	\bbm B_1 \\ D_{21} \ebm \right\| 
		$$
	In the state feedback setting this objective simplifies to
		$$
			\underset{K}{\inf}~ \|  \left[ \begin{array}{cc} C_1 & D_{12} \end{array} \right]\mathcal{F}\big(\tilde{P};K\big) B_1 \|.$$
	\item 
		The set of acheivable $\mathcal{F}\big(\tilde{P};K\big)$ is simple to characterize:
		\begin{multline}
			\Phi = 
			\bbm \Phi_{xx} & \Phi_{xy} \\ \Phi_{ux} & \Phi_{uy} \ebm = \mathcal{F}\big(\tilde{P};K\big) ~\text{for some $K$} 	\\
			\Leftrightarrow 
			\left\{
			\begin{aligned}
                        			&\bbm sI\sm A &\sm B_2 \ebm 
                        			\bbm \Phi_{xx} & \Phi_{xy} \\ \Phi_{ux} & \Phi_{uy} \ebm = \bbm I & 0 \ebm 		\\[3pt]
                        			&\bbm \Phi_{xx} & \Phi_{xy} \\ \Phi_{ux} & \Phi_{uy} \ebm  
            				\bbm sI\sm A \\ \sm C_2 \ebm 
            				= \bbm I \\ 0 \ebm   
			\\[3pt]
			&~\Phi_{xx}, \Phi_{ux}, \Phi_{xy} \text{ strictly proper}
			\end{aligned}	\right. 
		  \label{CL_Constraints.eq}
		\end{multline}
		This reduces in the state feedback setting:
\begin{multline}
			\Phi = 
			 \left[ \begin{array}{c} \Phix \\ \Phiu \end{array} \right] =\mathcal{F}\big(\tilde{P};K\big) ~\text{for some $K$} 	\\
			\Leftrightarrow 
			\left\{
			\begin{aligned} 
                        			&\bbm sI\sm A &\sm B_2 \ebm 
                        			\bbm \Phix  \\ \Phiu  \ebm = I 		\\[2pt] 
			& ~\Phix, \Phiu \text{ strictly proper}
			\end{aligned} \right. 
		  \label{CL_Constraints_state.eq}
		\end{multline}
where $\Phix$ and $\Phiu$ denote the closed-loop transfer functions from $B_1w$ to $x$ and $u$ respectively.
		\end{itemize} 
		
		Equipped with this terminology, we are able to define a third notion of structure.	
		\begin{defn}  \label{def:CL}
A controller $K$ is {\em closed-loop-transfer-function structured (CLTF-structured)} for plant $P$ with respect to graph $\cA$ 
	if each block entry of the resulting closed loop $\Phi={\cal F}(\Ptilde;K)$, partitioned as $\Phi  = \lba{cc} \Phi_{xx}& \Phi_{xy}\\ \Phi_{ux}&\Phi_{uy}\ear$ (output feedback) or as $\Phi = \left[ \begin{array}{c} \Phix \\ \Phiu\end{array} \right]$ (state feedback), is TF-structured with respect to $\cA$. When $P$ and $\cA$ are clear from context, we simply say $K$ is CLTF-structured. \end{defn}

\section{Controller Locality through Structured-Realizations \& Convex Relaxations} \label{sec:convexRelaxation}

\begin{figure}[t!]
\setlength\belowcaptionskip{-1\baselineskip}
	\begin{centering}
	\includegraphics[width=89mm]{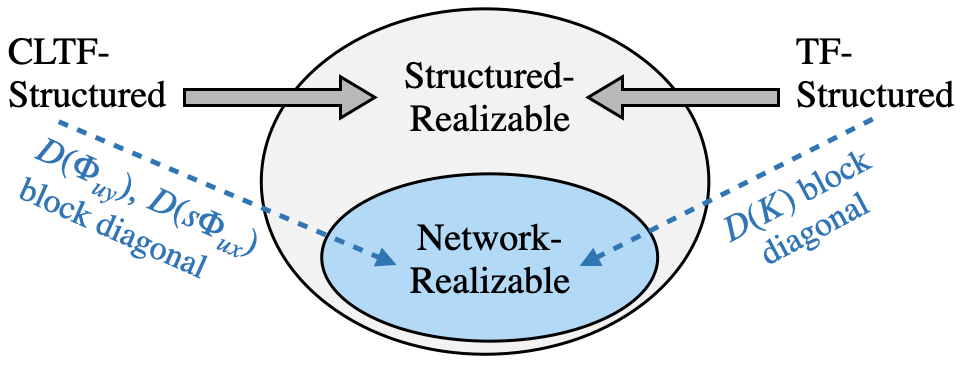}
 	 \caption{\footnotesize TF-structure and CLTF-structure are each sufficient conditions to ensure 
	 	controller structured-realizability. 
		Combined with added assumptions, these notions can also ensure controller network realizability.
		The notation $D(\Phi_{uy})$ refers to the direct feedthrough term of the transfer function $\Phi_{uy}$ and similarly for $D(s\Phi_{ux}(s))$ and $D(K)$. 
	 }
	 \label{fig:networked_and_structured_realizability}
	 \end{centering}
\end{figure}

In this section, we highlight that physical implementation of a distributed controller corresponds to a (typically non-minimal) structured realization. We prove that CLTF-structured controllers admit structured controller realizations (Thm.~\ref{thm:TF_CLTF_implySS}) and thus {\em structured closed-loop design is a convex relaxation of the structured controller realization design problem} (Cor.~\ref{cor:SLS_upperBound}).
The results of this section are summarized in Figure~\ref{fig:networked_and_structured_realizability}.

		An $\cA$-structured-realizable controller $u = Ky$ can be represented with 
	differential equations 
	\begin{subequations} \begin{align}
		\dot{\xi}_i(t) &= \sum_{\cA_{ij} \ne 0}A_{ij} \xi_j(t) + \sum_{\cA_{ij} \ne 0} B_{ij} y_{j} \label{eq:controllerState}\\
		u_i(t) & =  \sum_{\cA_{ij} \ne 0}C_{ij} \xi_j(t) + \sum_{\cA_{ij} \ne 0} D_{ij} y_{j}, \label{eq:controllerOutput},
	\end{align}\end{subequations}
	 using only information from sites within a local neighborhood (according to $\cA$) to compute a given local output, where $\xi_i, y_i, u_i$ are the sub-controller state, measured output and local control action respectively at spatial site $i$. 
In the case of network-realizability, the output equation \eqref{eq:controllerOutput} is further restricted to the form:
	\be \label{eq:networkControllerOutput}
		u_i(t) =  C_{ii} \xi_i(t) + D_{ii} y_i(t),
	\ee
which is completely localized in the sense that each local control action $u_i$ is computed using only the $i^{\rm th}$ sub-controller state and $i^{\rm th}$ measurement\footnote{The notion of network-realizability in \cite{vamsi2015optimal, rantzer} is slightly different than our Definition~\ref{def:structuredrealizable}, in that $A,C$ are constrained to be in $\mathcal{S}(\cA)$ while $B,D$ are constrained to be block diagonal. Under either choice of definition the set of network-realizable systems is convex and satisfies the mentioned closure properties. Here we restrict $C$ and $D$ to be block diagonal, rather than $B$ and $D$, as this is more natural when one considers e.g. discrete-time systems. }.

It is important to provide tractable methods for obtaining structured/network-controller realizations (and in particular admissible ones), as we see that these are utilized to physically implement a distributed controller. Mathematically, this problem of interest is represented by \eqref{eq:cl_opt_const} with structural constraint \eqref{eq:struct_require} given by  
	\begin{enumerate} [label=(\roman*)]
		\item $K$ has an admissible $\cA$-structured-realization, or
		\item $K$ has an admissible $\cA$-network-realization
	\end{enumerate}

The set of network-realizable systems was originally introduced in \cite{vamsi2015optimal}, and it was demonstrated in \cite{rantzer} that this set is convex and closed under  parallel, cascade and feedback interconnections. 
A characterization of the set of stable, admissible, network realizable systems was provided in \cite{vamsi2015optimal}, but an analogous characterization in the unstable setting has yet to be developed. Characterization of this set is difficult, as there is no restriction on the number of states in the realizations. 
Thus, controller design subject to constraints (i) or (ii) remains an open problem.

 To that end, {\em the main result of this section provides convex relaxations of the optimal structured- or network-realizable controller design problem.} We begin by providing sufficient conditions for structured- or network-realizability of a controller. 
 To simplify notation,  let $D(H)$ denote the direct feedthrough term of a transfer matrix $H$, i.e. $D(H) = \lim_{s \rightarrow \infty} H(s)$. 
 \begin{thm} \label{thm:TF_CLTF_implySS}
	Let $K$ be an LTI controller for the plant $P$. 
	Then the following hold: 
	\begin{enumerate}[label=(\alph*)]
		\item If $K$ is TF-structured, then $K$ is structured-realizable.  $K$ is additionally network-realizable if and only if $D(K)$ is block diagonal.
		\item If $K$ is CLTF-structured controller for $P$, then $K$ is structured-realizable. 
		If additionally $\Phi = \F(\tilde{P};K)$ is stable, $K$ has an  
		admissible structured realization. 
		Moreover $K$ has an admissible network realization under the following conditions:
		\begin{enumerate}[label=(\roman*)]
			\item {\it (State Feedback)} if and only if $D \left( s \Phiu \right)$ is block diagonal,
			\item {\it (Output Feedback)} if both $D \left( s\Phiu\right)$ and $D(\Phi_{uy})$ are block diagonal.
		\end{enumerate}	
		\end{enumerate}
		In other words, any TF-structured or CLTF-structured controller admits a (typically non-minimal) structured state-space realization. This result is illustrated in Figure~\ref{fig:networked_and_structured_realizability}. 
	\end{thm}

We remark that results similar to statement (a) of Thoerem~\ref{thm:TF_CLTF_implySS} have been presented in e.g. \cite{vamsi2015optimal}.  Nonetheless, an explicit proof of (a) is provided in Appendix~\ref{app:TFimpliesSS} for clarity and for use in proving subsequent results.

Statement (b) of Thoerem~\ref{thm:TF_CLTF_implySS}  provides the first formal proof to our knowledge that CLTF-structured controller design leads to structured controller realizations, even in the continuous time IIR setting. 
Closed-loop design in the context of SLS has largely been described as a convex method for structured controller ``implementation” design. However, this notion of implementation has been defined through a block diagram of interconnections of transfer functions in \cite{wang2019system} and related works, and thus is left somewhat vague. 
In this section, we aim to clarify this notion of implementation using the concepts of structured realizability and network realizability. 

Structured minimal realizations were derived for the special case of discrete time, FIR SLS controllers (i.e. an analogue of statement (b) was proven) in \cite{anderson2017structured}; the techniques utilized in \cite{anderson2017structured} are not generalizable to IIR settings, and indeed IIR filters may allow for realizations with lower state dimension (less memory) than their FIR approximations. Thus Theorem~\ref{thm:TF_CLTF_implySS} (b) is a contribution to the SLS and distributed control literature. 
To prove this result, we leverage the controller ``implementation" suggested by \cite{wang2019system} (with a modification to the continuous time setting). 
This implementation depicts the second key idea of the SLS framework which is summarized next.

\subsection{CLTF-Structured Controller Block Diagram}	
\begin{figure}[h]
	\centering 
		\begin{subfigure}{0.43\textwidth}
		\setlength\belowcaptionskip{-.5\baselineskip}
		\centering
			\includegraphics[width=\textwidth]{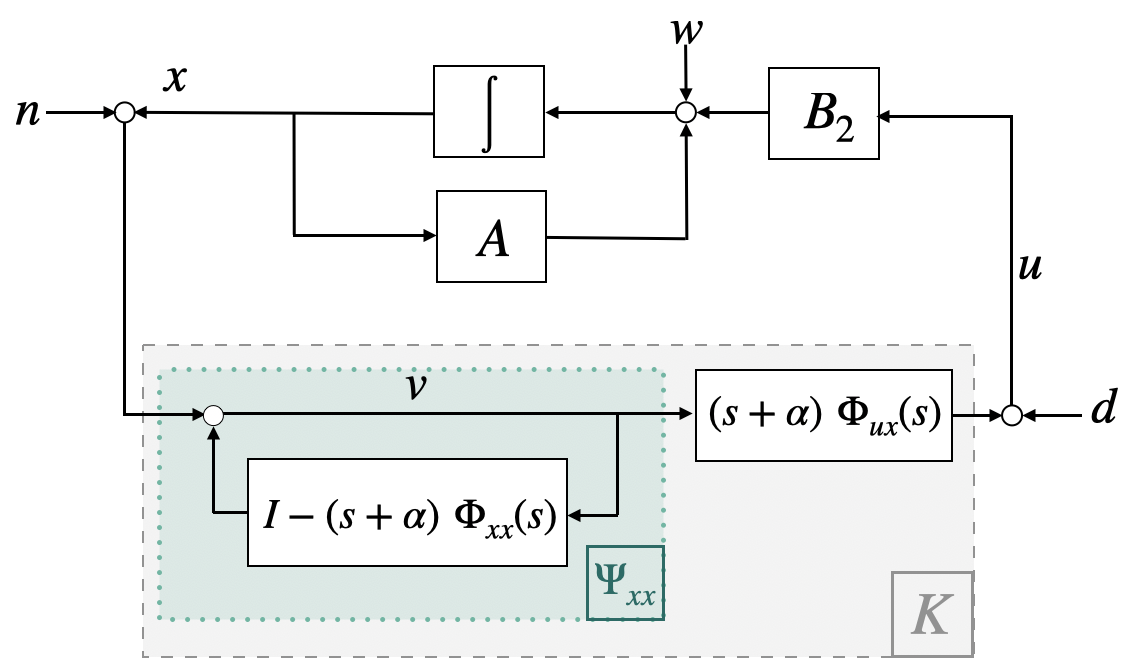}
			\subcaption{\footnotesize Feedback interconnection of plant $P$ with the LTI state feedback controller $K$ that results in closed-loop mappings $\Phi = [ \Phix ~ \Phiu]^T$. This block diagram implementation of $K$ (depicted by shaded grey box) preserves any structure imposed on $\Phi$. 
		For any choice of ${\rm Re}(\alpha) >0 $, this implementation does not introduce any unbounded internal signals (Lem.~\ref{lem:implementation}).}
			\label{fig:sls_controller}
		\end{subfigure}
			 
		\bigskip

	\centering 
		\begin{subfigure}{0.43\textwidth}
		\centering
			
			\includegraphics[width=\textwidth]{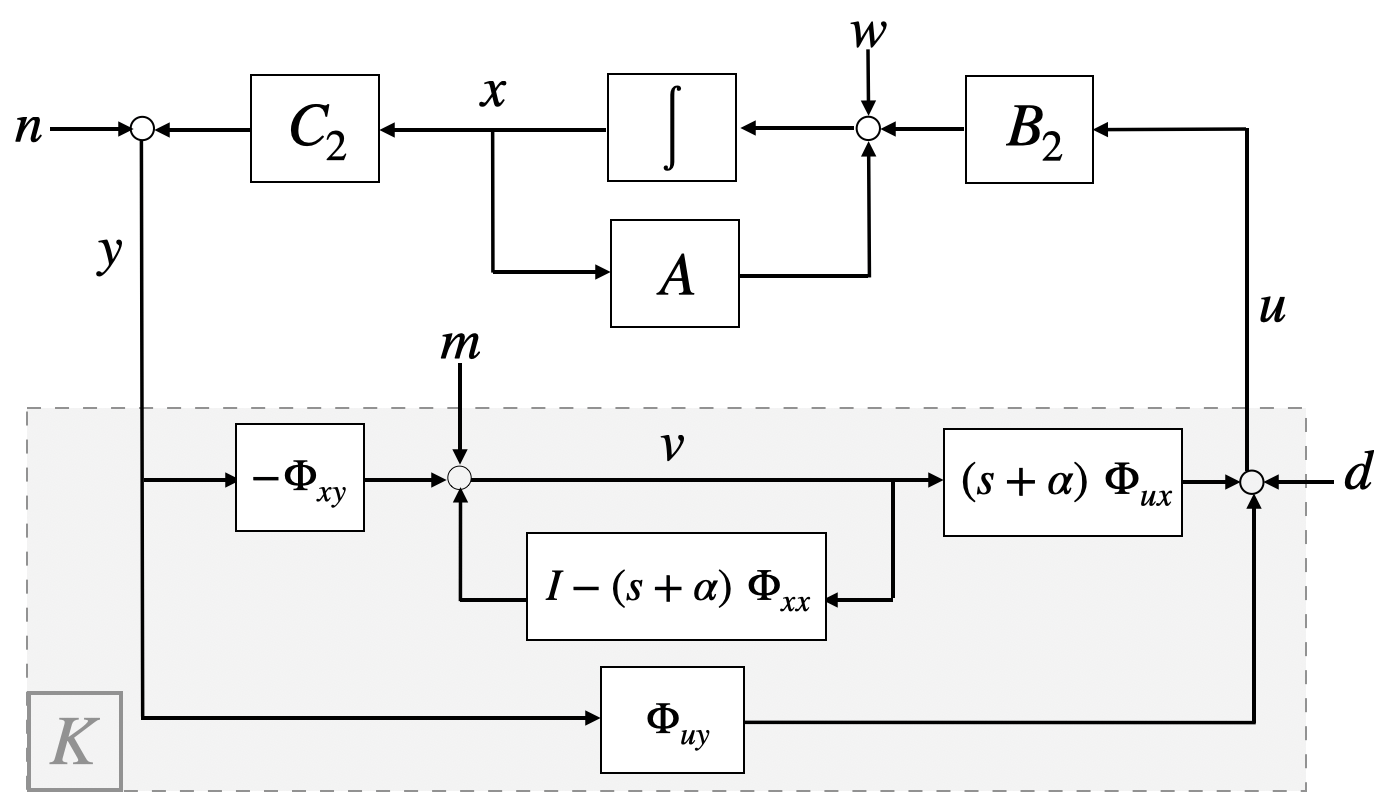}
			\subcaption{\footnotesize Feedback interconnection of plant $P$ with LTI output feedback controller $u = Ky$ that results in the closed-loop mappings $ \Phi_{xx} ,\Phi_{xy} ,\Phi_{ux} ,\Phi_{uy}$. 
			The chosen block diagram implementation of this controller $K$ (shaded grey box) preserves any structured imposed on the resulting closed-loops. For any choice of ${\rm Re}(\alpha) > 0$, this implementation does not introduce any unbounded internal signals (Lem.~\ref{lem:implementation}).}
		\label{fig:sys_and_outputFB}
		\end{subfigure}
		\caption{\footnotesize The second key idea of SLS is to implement the controller $K$ using components of the closed
		loop maps $\Phi$. }
		\label{fig:block_diagrams}
	\end{figure}

The second key idea of SLS is to construct an ``implementation" of $K$ from components of $\Phi$ that inherits the structure imposed on $\Phi$. In particular, the output feedback controller $u = Ky$ can be recovered from the components of the resulting mapping $\Phi$ as 
	\be \label{eq:K_output}
		K = \Phi_{uy} - \Phi_{ux}\Phi_{xx}^{-1} \Phi_{xy},
	\ee 
and ``implemented" in the continuous-time setting using the components of $\Phi$ as
	\be \ba \label{eq:SLS_outputImplementation}
		v & =  - \Phi_{xy} y +  \left( I - (s+\alpha) \Phi_{xx}\right) v \\
		u & = (s+\alpha) \Phi_{ux} v + \Phi_{uy} y,
	\ea \ee
with $\alpha$ a scalar-valued parameter (see Figure~\ref{fig:sys_and_outputFB}). Similarly the state feedback controller $u = Kx$  can be recovered from the resulting closed-loop $\Phi$ as  
	\be \label{eq:K_state}
		K= \Phiu \Phix^{-1}, 
	\ee
and ``implemented" in the continuous time setting using the components of $\Phi$ as 
		\begin{equation} \begin{aligned} \label{eq:SLSimplementation}
			v &= x + (I - (s+\alpha) \Phix(s)) v\\
			u &= (s+\alpha) \Phiu(s) v,
		\end{aligned} \end{equation}
(see Figure~\ref{fig:sls_controller}). Of course this choice of block diagram to preserve structure is not unique; alternative choices have been presented in e.g. \cite{9147953, li2020}. To see that these block diagrams (Figure~\ref{fig:block_diagrams}) are well-posed, note that strict-properness of $\Phix, \Phiu$ implies strict properness of 
	\be 
		I - (s+\alpha) \Phix \overset{(1)}{=} -(A+ \alpha I) \Phix - B_2 \Phiu,
	\ee
where equality $\overset{(1)}{=}$ follows from \eqref{CL_Constraints_state.eq}. We demonstrate that these implementations do not introduce any unbounded internal signals when we restrict to  ${\rm Re}(\alpha) >0$ and use this to prove Theorem~\ref{thm:TF_CLTF_implySS} (b) in Appendix \ref{app:SLSimpliesDist}.

\begin{rem} In addition to demonstrating that CLTF-structure implies structured controller realizations, 
Theorem~\ref{thm:TF_CLTF_implySS} also unifies the two seemingly distinct problems of ``sparse controller" design and ``sparse closed-loop" design as two methods of designing a controller with a distributed implementation (see Figure~\ref{fig:networked_and_structured_realizability}).
\end{rem}
		
We mention that the converse of Theorem~\ref{thm:TF_CLTF_implySS} does not hold, as stated in the following proposition. 

\begin{prop} \label{prop:notTF_CLTF}
 There exist structured-realizable (and network-realizable) systems $K$ which are not TF-structured and  which are not CLTF-structured. 
	\end{prop}
\begin{IEEEproof} See  Appendix~\ref{app:notTF_CLTF}.\end{IEEEproof}

\subsection{CLTF-Structured Design as Convex Relaxation of Structured-Realization Design}
The following is an immediate corollary of Theorem~\ref{thm:TF_CLTF_implySS}. 

	\begin{cor} \label{cor:SLS_upperBound}
		The optimal structured-realizable controller design problem is upper bounded by the CLTF-structured controller design problem:
			\begin{subequations}
				\begin{align}
			&\begin{array}{l}\inf_{K } ~\| \mathcal{F}(P;K) \|  \\
			~ {\rm s.t.}~   K\text{ has $\cA$-structured realization}\end{array} \label{eq:opt_structured} \\
			\le ~ & \begin{array}{l} 
			\inf_{K} ~\| \mathcal{F}(P;K) \|  \\
			~ {\rm s.t.}~K ~\text{CLTF-structured w.r.t. }\cA
			\end{array} \label{eq:CLTF_opt}
			\end{align}
			\label{eq:sr_bound}
			\end{subequations}	 
With additional constraints, the admissible network-realizable controller design problem is also bounded:
			\begin{subequations}
				\begin{align}
			&\begin{array}{l}\inf_{K } ~\| \mathcal{F}(P;K) \|  \\
			~ {\rm s.t.}~~   K\text{ has admissible $\cA$-network realization}\end{array} \label{eq:opt_network} \\
			  &  \begin{array}{l} \le ~
			\inf_{K} ~\| \mathcal{F}(P;K) \|  \\
			~~~~~ {\rm s.t.}~~K ~\text{CLTF-structured w.r.t. }\cA, \\
			~~~~~~~~~~~ \Phi~\text{stable}, \\
			~~~~~~~~~~~ D(s \Phi_{ux}), D(\Phi_{uy})\text{ block diagonal}
			\end{array} \label{eq:CLTF_opt_net}
			\end{align}
			\label{eq:anr_bound}
			\end{subequations}	 
	\end{cor}

	Note that the optimal TF-structured controller design problem also upper bounds \eqref{eq:opt_structured} but is non-convex in general. The CLTF-structured controller design problem \eqref{eq:CLTF_opt} is always convex and remains convex with the added convex constraints in \eqref{eq:CLTF_opt_net}. Thus, {\em CLTF-structured design provides a convex method to bound the structured or networked-realizable controller design problem.} When \eqref{eq:CLTF_opt} has a stable solution $\Phi$, it provides a stabilizing, admissible, structured-realizable controller for the plant $P$. In the case that \eqref{eq:CLTF_opt} is infeasible however, we gain no information about the existence or performance of a structured-realizable controller.

\section{Relative Feedback as a Design Constraint} \label{sec:relative_feedback}
In this section, we move away from locality and consider another structural constraint of interest: relative feedback. We demonstrate relative feedback requirements can be written as a convex constraint on the controller transfer matrix (Thm.~\ref{lem:relativeK}-a) and (under certain settings) as a convex constraint on the closed-loop (Thm.~\ref{lem:relative}), allowing for efficient use with SLS.  We also provide a characterization of the communication structure of relative feedback controllers (Thm.~\ref{lem:relative}-b).

A controller (or general dynamical system) is \emph{relative} if its outputs can be computed using only relative differences of inputs; this is formalized in the following definition.

	\begin{defn} \label{defn:relative}
            	Consider signals $u$ and $y$ partitioned into sub-signals as in~\req{u_partition}, 
            	and a transfer function matrix $K$ partitioned conformably with $u$ and $y$ as follows 
		\be
			\bbm u_1 \\ \vdots \\ u_\ssM \ebm = 
			\bbm K_{11} & \cdots & K_{1\ssN} \\
					\vdots &  & \vdots \\ 
				K_{\ssM 1} & \cdots & K_{\ssM\ssN} 	\ebm 
					\bbm y_1 \\ \vdots \\ y_\ssN \ebm , 
		  \label{eq:K_partitions} 
		\ee
		where $\lcb y_i \rcb$ are vector signals all of the same dimension. 
		The $m$'th block-row of 
            	$K$ is called  \emph{relative} if there exists a collection 
		$\lcb\cK^{(m)}_{ij}\rcb_{1\leq i <j\leq \ssN}$ of transfer function matrices such that 
            	\be 
            		u_m = 
				\sum_{j=1}^\ssN K_{mj} ~y_j 
				= 
				\sum_{1\leq i < j \leq \ssN}  \cK^{(m)}_{ij} ~  (y_i - y_j), 
            	 \label{relative_def.eq}
            	\ee
		i.e., if  $u_m$ is obtained from only differences of inputs. 
		The entire matrix $K$ is called {\em relative} if each block-row is relative. 
	 \end{defn}

	To characterize such relative transfer function matrices, we introduce the following vector or matrix 
	\[
		\mathbb{1} := \bbm 1 \\ \vdots \\ 1 \ebm  ~
		\mbox{\begin{minipage}{.08\textwidth} if each $y_i$ scalar \end{minipage}},~~
		\mathbb{1} :=  \bbm I \\ \vdots \\ I \ebm~
		\mbox{\begin{minipage}{.18\textwidth} $I$ an identity matrix of same dimension as $y_i$. \end{minipage}}
	\]
	\begin{thm} \label{lem:relativeK}
	Let $K$ be a transfer matrix partitioned as in~(\ref{eq:K_partitions}), 	
	and let $\lcb \cA^{(m)}  \rcb $ be a collection of undirected graphs.
		\begin{enumerate} [label=(\alph*)]
		\item The $m$'th block-row $K^{(m)}$ of $K$ is {relative} 
			 if and only if
		\be \label{eq:K1}
			K^{(m)}(s) ~\mathbb{1} = 0.
		\ee 
		Thus the entire matrix $K$ is relative iff $K(s) \mathbb{1}=0$. 
		\item If $K^{(m)}(s) \mathbb{1} = 0$, then there
		exists transfer function matrices $\lcb \cK^{(m)}_{ij}\rcb$ such that $K$ can be 
		implemented as in~\eqref{relative_def.eq} 
		with the same sparsity structure as $\cA^{(m)}$, i.e. 
	\[
		 \cA^{(m)}_{ij}=0 ~~~\Rightarrow~~~\cK^{(m)}_{ij} = 0 
	\]
	if and only if $\cA^{(m)}$ is connected. 
	\end{enumerate}
	\end{thm}

\begin{IEEEproof} See Appendix~\ref{lem:relativeK}. \end{IEEEproof}

\begin{figure}[t!]
\setlength\belowcaptionskip{-1\baselineskip}
	\begin{centering}
	\includegraphics[width=70mm]{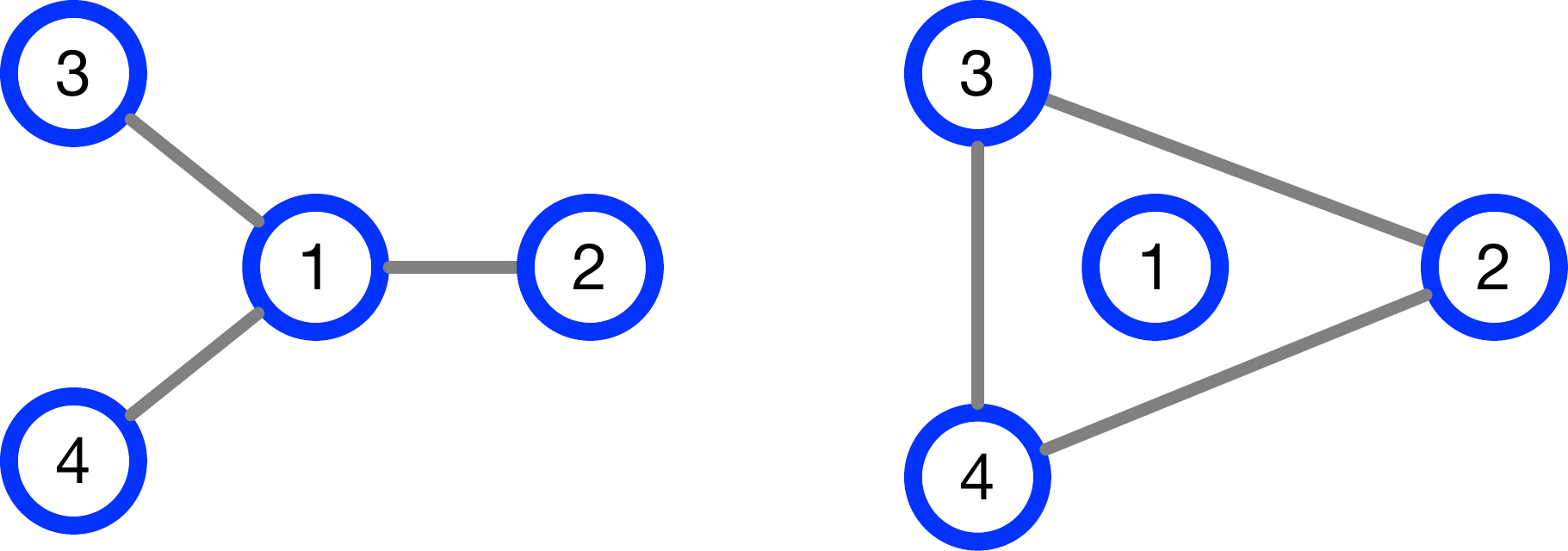}
 	 \caption{\footnotesize An example demonstrating the difference between locality (left) and relative feedback (right) graphs. 
	 	The locality graph implies that the controller for node 1 can use measurements from nodes 2-4. The relative feedback 
		graph implies that the the controller for node 1 can only use the differences $y_2-y_3$, $y_3-y_4$ and $y_4-y_2$. 
	 }
	 \label{fig:locality_relative_graphs}
	 \end{centering}
\end{figure}

Note the similarity of these conditions to those of ``right-stochastic matrices''. 
A version of the requirement \eqref{eq:K1} has been utilized in earlier works, e.g. \cite{bamieh2012coherence}. However, the proof of this result along with the characterization of structures (statement (b)) is a novel contribution. 
Each edge $(i,j)\in\cA^{(m)}$ specifies which differences of subsets of the measurement $y$ are available to $u_m$. We note that this  graph is unrelated to the graph that specifies locality - relative feedback and locality are mutually independent concepts. 
The example  in Figure~\ref{fig:locality_relative_graphs} illustrates this point. 

We will need a further refinement of the concept of relative systems which we state next for the entire matrix $K$, with the obvious parallel statements applying to the cases of single rows. 
Under the relative feedback constraint $K \one = 0$, the controller $K$ may use differences of \emph{any} inputs to compute its output. Thus, any two inputs $y_i$ and $y_j$ must be \emph{compatible}, i.e. their difference $(y_i - y_j)$ must be physically meaningful. In many applications though, all inputs are not compatible. We specifically consider the case when the inputs can be partitioned into two distinct categories as 
	\be
		y= \lba{cccccc} y_1 & \cdots & y_n  ~\vline ~y_{n+1} & \cdots & y_\ssN \ear^\ssT 
				=: \lba{c} y^{(1)} \\ y^{(2)}\ear
	\ee
where $y_1, ..., y_n$ are compatible, and $y_{n+1}, ..., y_N$ are compatible. For example, the inputs $y^{(1)}$ may correspond to individual vehicle position measurements in a platoon while inputs $y^{(2)}$ correspond to individual vehicle velocity. In this case, the controller may use difference of positions of any two vehicles and differences of velocities of any two vehicles, but not e.g. the difference between position of one vehicle and velocity of another. Partitioning the controller conformably as 
	\be
		u = \bbm K_1 & K_2 \ebm  \bbm y^{(1)} \\ y^{(2)}\ebm,
	\ee
the constraint of interest is mathematically of the form 
	\be\label{eq:partitionedKconstraint}
		\bbm K_1 & K_2 \ebm \bbm \mathbb{1} \\ 0 \ebm = 0 
		\hspace{1em} \mbox{and} \hspace{1em}
		 \bbm K_1 & K_2 \ebm \bbm 0 \\ \mathbb{1} \ebm = 0.
	\ee
Note that \eqref{eq:partitionedKconstraint} is stricter than the constraint $K \mathbb{1} = 0$, but remains convex. Obvious generalizations to convex constraints in the case of larger number of distinct classes of inputs can be made. 
\color{black}

\subsection{Relative Feedback \& Realizations}

We next remark on what Theorem~\ref{lem:relativeK} tells us about state-space realizations of relative controllers. 
	\begin{defn}
		A realization of a controller $u = K y$:
			\be \ba \label{eq:controlEq}
				\dot{\xi} &= A \xi + B y \\
				u & = C \xi + D y 
			\ea \ee
		is  \emph{relative} if computation of the state update and output equations \eqref{eq:controlEq} only require access to differences of inputs $(y_i - y_j)$. Equivalently  the $B$ and $D$ matrices are relative, i.e. $B \mathbb{1} = 0$ and $D \mathbb{1} = 0 $. 
	\end{defn}
	
If $K$ has a relative realization, then 	$$
		K(s) \mathbb{1} = C(sI - A)^{-1} B \mathbb{1} + D \mathbb{1} = 0, 
	$$
so that $K$ is a relative according to Theorem~\ref{lem:relativeK}. The following proposition provides a partial converse to this statement; the proof is provided in Appendix~\ref{app:relative_realization}.
\begin{prop} \label{prop:relative_realization}
	Let $K$ be a relative transfer function, i.e. $K \mathbb{1} = 0$, with realization $K = \lba{c|c} A & B \\ \hline C & D \ear$. If $(C,A)$ is observable then this realization must be relative, i.e. $B \mathbb{1} =0,$ and $D\mathbb {1} = 0$
\end{prop}

\subsection{Closed-Loop Parameterization of Relative Feedback} \label{sec:CLrelative}
We next demonstrate that, in certain settings, a relative feedback constraint on the controller can be written as a convex constraint on the closed loops.

\begin{thm} \label{lem:relative}
Let $u = Kx$ be the LTI (not necessarily static) state feedback controller for plant $P$ with dynamics
	\be
		\dot{x} = A x + B_1 w + B_2 u,
	\ee
and assume $B_2$ is full rank.
		\begin{enumerate}  [label=(\alph*)]
\item Assume $A$ relative, i.e. $A \mathbb{1} = 0$. Then $K$ is relative if and only if the resulting closed-loop transfer function $\Phiu$ is relative, i.e. 
		$
			\Phiu(s) \mathbb{1} = 0.
		$ 
	\item Assume $A$ and $B_2$ can be block partitioned as 
		\be \label{eq:ABpartitioned}
			A = \lba{cc} A_1 & I \\ A_2 & A_3 \ear, ~ B_2 = \lba{c} 0 \\ \overline{B} \ear
		\ee
	where $A_i \mathbb{1} = 0$ for $ i= 1,2,3$. Then $K$ satisfies constraint \eqref{eq:partitionedKconstraint} if and only if the resulting closed-loop transfer function $\Phiu$ satisfies 
		\be \label{eq:PhiuPartitionedRelative}
			 \Phiu(s) \lba{c} \mathbb{1} \\ 0 \ear = 0, ~ \Phiu(s)  \lba{c} 0 \\ \mathbb{1} \ear = 0.
		\ee
	\end{enumerate} 
\end{thm}

\begin{IEEEproof}
See Appendix~\ref{relativeSLS.appendix}.
\end{IEEEproof}

The assumptions of Thoerem~\ref{lem:relative} (a) hold for systems that obey underlying conservation laws, such as diffusion or consensus. Systems that satisfy the assumptions of Thoerem~\ref{lem:relative} (b) include the vehicular platoon problem \cite{bamieh2012coherence}, the wave equation \cite{curtain2020introduction}, and the swing equations of electrical power networks \cite{dorfler2014sparsity}. 
\color{black}

\section{Performance Gap} \label{sec:gap}
Problems \eqref{eq:opt_structured} and \eqref{eq:opt_network} are non-convex, with no known tractable solutions. \eqref{eq:CLTF_opt} and \eqref{eq:CLTF_opt_net} provide tractable suboptimal solutions to these problems.  
An important step toward assessing the usefulness of these bounds is to quantify the associated performance gap. The main result of this section (Thm.~\ref{thm:infeasibility}) provides a partial answer: {\em we construct a class of examples (with the additional convex constraint of relative feedback) for which the performance gap between \eqref{eq:opt_structured} and \eqref{eq:CLTF_opt} is infinite.}

Consider the $\h_2$ design problem for the consensus of first-order subsystems presented in Section~\ref{sec:consensus}. 
We now use the framework of the previous section to formalize it. 
Let the graph $\ocA$ be given by a ring of $N$ nodes with nearest neighbor edges, i.e. 
	\be
		\ocA_{ij} = \begin{cases}
		1, ~ |i -j|~{\rm mod}~N = 1 \\ 0, ~ {\rm else}.
		\end{cases}
	\ee
Then an upper bound on the structured-realizable relative controller design problem:
	\begin{subequations} \begin{align}
	 \label{eq:opt_consensus_cost}
		&\inf_K~~\left\| \fdbk{P}{K} = \left[ \begin{array}{cc} C & 0 \\ 0 & \gamma I \end{array} \right] \left[ \begin{array}{c} \Phix \\ \Phiu\end{array} \right] \right\|_{\h_2}		\\
		 &~{\rm s.t.} ~ ~K ~\text{relative }\\
		 &~~~~~~~K \text{ is }\ocA^{(b)}\text{-structured-realizable},
	\end{align}
	 \label{eq:opt_consensus1}
	\end{subequations}
is provided by the following example.
\begin{exmp}
	The controller presented in Eq.~\eqref{eq:standard_consensus} (which results in the standard consensus algorithm) is relative, $\ocA$-structured-realizable and TF-structured w.r.t. $\ocA$. When $k_- =  k_+ = 1$, \eqref{eq:standard_consensus} is equivalently represented by the static controller transfer matrix
		\begin{equation} \label{eq:Ks}
		K_s :=   \left[ \begin{array}{cccccc} -2 & 1 & 0 & \cdots & 1 \\ 1 & -2 & 1 & \cdots  & 0\\& \ddots & \ddots & \ddots  \\ 1 & 0 & 0 & \cdots  & -2\end{array} \right].
	\end{equation} 
Thus
	\begin{equation*} \begin{aligned}
 	 	&\| \mathcal{F}(P;K_s) \|_{\mathcal{H}_2}^2 = {\rm tr}  \int_{0}^{\infty} e^{K_s^*t} \left[ \begin{array}{cc} C^{\rm ave.} \\ \gamma K_s \end{array} \right]^*\left[ \begin{array}{cc} C^{\rm ave.} \\ \gamma K_s \end{array} \right] e^{K_st} dt \\
		 &~~~~~~=\frac{N-1}{4} + \gamma^2 \sum_{n=0}^{N-1} \cos^2\left(\frac{2 \pi n}{N}\right) < \infty
	\end{aligned} \end{equation*}
is an upper bound for \eqref{eq:opt_consensus1} with $b=1$ and $C = C^{\rm ave.}.$
\end{exmp}
	
We build on this example to bound the admissible, network-realizable controller design problem by constructing a strictly proper approximation of $K_s$.
\begin{prop} \label{prop:Ka}
As $a \rightarrow -\infty$ the closed-loop performance of  \eqref{eq:plantState} in feedback with 
	\be \label{eq:Ka}
			K_a:= \left[ \begin{array}{c|c} aI & K_s \\ \hline aI & 0 \end{array} \right],
		\ee
converges in $\h_2$ norm to the performance of \eqref{eq:plantState} in feedback with  $K_s$, i.e.
 	\be
		\| \F(P; K_s) - \F(P; K_a)\|_{\h_2} \underset{a \rightarrow -\infty}{\longrightarrow} = 0
	\ee
\end{prop}
\begin{IEEEproof}
		See Appendix \ref{app:propKa}. 
	\end{IEEEproof}

$K_a$ is $\ocA$-network-realizable and admissible, so that the following upper bound holds:
	\begin{subequations} 
	\begin{align}
	&\| \F(P; K_a)\|_{\h_2}~~ \ge \\ 
		&  \label{eq:opt_consensus2}
		\left. \begin{array}{l}
		\inf_K~ \| \fdbk{P}{K} \|_{\h_2}						\\
		 {\rm s.t.} ~~K ~\text{relative }\\
		 ~~~~~~K \text{ has admissible }\ocA^{(b)}\text{-network realization}
	\end{array} \right\}
	\end{align}
	\end{subequations}
when $b=1$ and $C = C^{\rm ave.}$.	
		
	$K_s$ and $K_a$ provide finite upper bounds for  \eqref{eq:opt_consensus1} and \eqref{eq:opt_consensus2}. By Corollary~\ref{cor:SLS_upperBound}, the CLTF-structured, relative $\h_2$ controller design problem  
 provides a more systematic, convex approach for obtaining upper bounds:
 	\begin{subequations}
	\begin{align}
			&\inf_K ~ ~\left\| \fdbk{P}{K} = \left[ \begin{array}{cc} C & 0 \\ 0 & \gamma I \end{array} \right] \left[ \begin{array}{c} \Phix \\ \Phiu\end{array} \right] \right\|_{\h_2} \label{cost} \\
			&~{\rm s.t.}~ ~K\text{ CLTF-structured w.r.t. }\ocA^{(b)} \label{const1} \\
			&~~~~~~~K ~\text{relative} \label{const2} \\
			&~~~~~~~\Phi~\text{stable},~ D(s\Phi_u)~\text{block diagonal} \label{const3}.
	\end{align}\label{eq:infeasible_exmp}\end{subequations}
With just constraints \eqref{const1}-\eqref{const2}, problem \eqref{eq:infeasible_exmp} bounds the structured-realizable design problem \eqref{eq:opt_consensus1}. With all constraints \eqref{const1}-\eqref{const3}, problem \eqref{eq:infeasible_exmp} bounds the admissible network-realizable design problem \eqref{eq:opt_consensus2}. 
These bounds are quantified in the following theorem.

\begin{thm} \label{thm:infeasibility}
Consider the CLTF-structured relative $\h_2$ controller design problem
\eqref{eq:infeasible_exmp} subject to the two constraints  \eqref{const1}-\eqref{const2} with $C = C^{\rm ave.}$, or more generally any $C$ satisfying conditions:
\begin{enumerate}[label=(\roman*)]
	\item $C$ is circulant, i.e. of the form 
		\be
			 C = \lba{cccc} c_0 & c_{N-1} & \cdots & c_1 \\ c_1 & c_0 & \cdots & c_2 \\ \vdots & & \ddots & \vdots \\c_{N-1} & c_{N-2} &\cdots & c_0  \ear, 
		\ee
	\item $C$ is relative.
		\end{enumerate}
The corresponding optimal cost \eqref{cost} is infinite whenever
	\be
		{\rm rank}(C) > 2b + 1,
	\ee
i.e. problem \eqref{cost}-\eqref{const2} is infeasible.
\end{thm}

Thus, there is an infinite performance gap between \eqref{eq:opt_consensus1} and \eqref{cost}-\eqref{const2}. With the additional constraint \eqref{const3} imposed, the optimal cost of will clearly remain infinite. Thus, the performance gap between \eqref{eq:opt_consensus2} and \eqref{cost}-\eqref{const3} is also infinite.

The physical interpretations of assumptions (i)-(ii) are as follows: Under assumption (i) the closed-loop cost is invariant to spatial shifts;
under assumption (ii) the closed-loop cost $\| \F (P;K) \|_{\h_2}$ depends only on differences in position e.g. $(x_i - x_j)$ and not on absolute position. A marginally stable mode at the origin of the closed-loop $\Phix$ (representing motion of the mean of the $x_i's$) will be undetectable under assumption (ii), but stability of the input-output mapping $\F(P;K)$ 
is required for the closed-loop cost to be finite.
In addition to $C^{\rm ave.}$, two other choices satisfying (i)-(ii) are measures of 
\begin{itemize}
		\item 
			$y_n = \left( C^{\rm l.e.} x \right)_n:=x_n - x_{n-1}$ ~~~(\emph{Local Error})
		
	\item 
			$y_n = \left( C^{\rm l.r.} x \right)_n:= x_n - x_{(n - N/2)}$~~~(\emph{Long Range Deviation})
\end{itemize}
Thus, infeasibility of \eqref{cost}-\eqref{const2} can occur for both {local} (e.g. $C^{\rm l.e.}$) and  {global} measures of consensus (e.g. $C^{\rm ave.}$ and $C^{\rm l.r.}$).

\begin{IEEEproof} (of Theorem~\ref{thm:infeasibility})
Using the parameterization of Theorem~\ref{lem:relative} along with parameterization \eqref{CL_Constraints_state.eq} we write \eqref{eq:infeasible_exmp} equivalently as
	\begin{equation}  \label{eq:SLSconsensus}
	\begin{array}{cl}
	 \underset{\Phiu, \Phix }{\inf}& \left\|\F(P;K) = \left[ \begin{array}{cc} C & 0 \\ 0&\gamma I  \end{array} \right] \left[ \begin{array}{c} \Phix \\ \Phiu  \end{array} \right]  \right\|_{\mathcal{H}_2}^2 \\
	\text{s.t.}& \left[ \begin{array}{cc} sI - 0 & -I \end{array} \right] \left[ \begin{array}{c} \Phix(s) \\ \Phiu(s) \end{array} \right] =I\\
		& \Phix, \Phiu \text{ strictly proper}\\	
		&  \Phix,\Phiu \text{ TF-structured w.r.t. }\ocA^{(b)}~\text{(locality)}\\
	& \Phiu \mathbb{1} = 0~~~~~~~~~~~~~~~~~~~\text{(relative feedback)}
	\end{array} 
	\end{equation}
Rearranging the affine subspace as
	$
	\Phix = \frac{1}{s} \left( I + \Phiu \right),
	$ %
we see that if $\Phiu$ is strictly proper and TF-structured w.r.t. $\ocA^{(b)}$, then $\Phix$ will be strictly proper and TF-structured w.r.t. $\ocA^{(b)}$ as well.
Then \eqref{eq:SLSconsensus} can be written equivalently as
	\begin{equation}  \label{eq:SLSconsensusPhiU}
	\begin{array}{cl}
	 \underset{\Phiu  }{\inf}& \left\| \F(P;K) = \left[ \begin{array}{cc} C & 0 \\ 0 & \gamma I\end{array} \right] \left[ \begin{array}{c}\frac{1}{s} \left( I - \Phiu \right) \\ \Phiu  \end{array} \right]  \right\|_{\mathcal{H}_2}^2\\
	\text{s.t.}&\Phiu \text{ strictly proper}\\
		& \Phiu \text{ TF-structured w.r.t. } \ocA^{(b)}~~~~~~~~~\text{(locality)}\\
	&\Phiu \mathbb{1} = 0 ~~~~~~~~~~~~~~~~~~~~~\text{(relative feedback)}
	\end{array}
	 \end{equation}

When the relative feedback requirement is removed, \eqref{eq:SLSconsensusPhiU} can be converted to a standard model-matching problem \cite{jensen2020Backstepping}, \cite{jensen2020} and solved with well-established techniques \cite{francis1987course}.
With the relative feedback constraint however, any $\Phiu$ in the constraint set of \eqref{eq:SLSconsensusPhiU} leads to an unstable $\F(P;K)$ and thus an infinite cost.  
Details of this argument are provided in Appendix~\ref{app:proofCompletion}.
\end{IEEEproof}

\begin{rem}
	Relative $K$ will have a minimal relative realization (by Proposition~\ref{prop:relative_realization}). However, it is unclear whether a relative and CLTF-structured $K$ will have a realization that is \emph{both} structured and relative. We do not impose the additional restriction of a relative structured realization in \eqref{eq:SLSconsensus}, noting that optimization over this smaller subset could only increase the size of the resulting performance gap.
\end{rem}

\subsection{Effects of Graph Structure} \label{sec:higher_dim}
In this section, we demonstrate that increasing spatial dimension (which can also be viewed as a proxy for graph connectivity) does not alter the infeasibility result of Theorem~\ref{thm:infeasibility}.
This result is motivated by the fact that static and first-order relative, localized controllers are able to regulate large-scale disturbances for the consensus problem in spatial dimension $d = 3$ but not in spatial dimension $d=1$ \cite{bamieh2012coherence}, \cite{tegling2019fundamental}. 
For the interested reader, this is stated and proved rigorously next, although the remainder of this section may be skipped without loss of continuity. 

Consider the problem of distributed consensus of $N^d$ $\first$-order subsystems on the undirected $d$-dimensional torus $\Z_N^d$:
	\begin{equation} \begin{aligned}  \label{eq:consensusHigherDim}
		\dot{x} &= u+ w,\\
		z &=\left[ \begin{array}{c} T_c \\ 0 \end{array} \right] x ~+~ \left[ \begin{array}{c} 0 \\  \gamma I \end{array} \right]u,
	\end{aligned} \end{equation}
where 
	$
		(T_c x)_n = x_n - \frac{1}{N^d} \sum_{m \in \Z_N^d} x_m.
	$
 corresponds to a deviation from average measure of consensus. 

For simplicity of exposition, we restrict our analysis to the {spatially-invariant} setting, e.g. the spatially-invariant controller $K$ is defined by convolution kernel $\{k_m(s)\}_{m \in \Z_N^d}$:
		\begin{equation} \begin{aligned}  \label{eq:K_higherDim}
			(KX)_n(s) &= (T_kX)_n(s) \\
			&:= \sum_{m \in \Z_N^d} k_m(s) X_{n-m}(s) = K (s) X(s),
		\end{aligned} \end{equation}
In feedback with $K$, 
the closed-loop mapping from disturbance $w$ to performance output $z$ is given by
	\be \label{eq:cl_higherDim}
		z = \mathcal{F}(P;K) w= \left[ \begin{array}{cc} T_c & 0 \\ 0 & \gamma I \end{array} \right] \left[ \begin{array}{c} \Phix \\ \Phiu \end{array} \right]w.
	\ee 
The mappings $\Phix$ and $\Phiu$ defined by \eqref{eq:cl_higherDim} are spatially-invariant systems \cite{jensen2018optimal}, i.e. of the same form as \eqref{eq:K_higherDim} with convolution kernels
 $\{(\Phix)_m(s)\}_{m \in \Z^d_N}$ and $\{(\Phiu)_m(s)\}_{m \in \Z^d_N}$ respectively.

To state the controller design problem of interest, we extend the necessary components to this higher-dimensional setting: 

\textit{1. (Performance Metric)} For spatially-invariant system $H$ with convolution kernel $\{h_m(s) \}_{m \in \Z_N^d}$ we define
	\be \label{eq:h2_higherDim}
		\| H \|_{\h_2}^2 :=  \sum_{m \in \Z_N^d} \|h_m \|_{\h_2}^2.
	\ee
We say if $H$ is stable if $\| H \|_{\h_2}^2 < \infty$. 

\textit{2. (Relative Measurements)} It is straightforward to show that a spatially-invariant system $K$ is \emph{relative} if and only if $K T_{\mathbb{1}} = 0$, where $T_{\mathbb{1}}$ denotes convolution with the array of all ones in $\Z_N^d$.

\textit{3. (Locality)}
Let $\cA$ denote the $d$-dimensional torus of $N^d$ nodes with edges between nearest neighbors, i.e. for two nodes defined by the multi-indices $i= (i_1, ..., i_d)$ and $j = (j_1, ..., j_d)$, 
	\be
		\cA_{ij} = 1 ~ \text{ iff } \max_{1 \le k \le d}| i_k - j_k ~({\rm mod~} N)| \le 1.
	\ee  
A spatially-invariant controller $K$ is \emph{CLTF-structured} for $P$ w.r.t $\cA^{(b)}$ if the convolution kernels of the closed-loops satisfy $(\Phiu)_n(s) = 0$, $(\Phix)_n(s) = 0$ for $\underset{1 \le j \le d}{\max} |n_j| > b$. 

Thus, the relative and CLTF-structured design problem in this higher-dimensional setting can be formally stated as:
		\begin{equation}  \label{eq:InfCLhigherDim}
		\begin{array}{cl}
			\underset{K}{ \inf}& \| \mathcal{F}(P;K) \|_{\h_2}^2  \\
			\text{s.t.}& K \text{ spatially-invariant,}\\
						& \text{$K$ CLTF-structured for $P$ w.r.t. }\cA^{(b)}  ~\text{(locality)}\\
			&K T_{\mathbb{1}} = 0 ~\text{(relative feedback)}
		\end{array}
		 \end{equation}

\begin{prop} \label{thm:higherDim}
	Problem \eqref{eq:InfCLhigherDim} is infeasible for any $1 \le b< \frac{N}{2}$. In other words, 
if $K$ is relative, spatially-invariant, and CLTF-structured for \eqref{eq:consensusHigherDim}, then
$\mathcal{F}(P;K)$ is not stable. 
\end{prop}

Further details and a proof of this proposition are available in \cite[Appendix 3.7.7]{jensen2020topics}.

\section{Discussion} \label{sec:discussion}


 In this section, we discuss three specific issues that are highlighted by the results of the present paper. These remain as  significant open questions for further research.

\subsection*{Alternate Closed-Loop Structures}
Physically, CLTF-structured controllers \emph{localize} the propagation of disturbances, e.g. if 
\be
	(\Phi_{\eta})_{ij}(s) = 0, ~~\eta \in \{xx, xy, ux, uy\}
	\label{eq:stringent} 
\ee 
then a disturbance entering at spatial site $j$ will not effect spatial site $i$ {\em for all time}. In contrast, structured-realizable controllers (that are not CLTF-structured) allow for disturbances to propagate to all spatial sites but with time lags that are proportional to the graph distance between sites. 
The disturbance attenuation requirement~(\ref{eq:stringent}) of CLTF-structured controllers appears to be a very stringent additional requirement to enforce over a structured-realizability requirement, as evidenced by the infinite performance gap of Theorem~\ref{thm:infeasibility}.  Thus, rather than imposing a sparsity structure on the closed-loop transfer functions, it may make more sense in certain problem settings to impose a less extreme structure. Indeed, other structural constraints on the closed-loops will still be convex. One option, inspired by \cite{vamsi2015optimal}, is a relative degree requirement on the entries of the closed-loop transfer functions. This may be advantageous over the sparsity requirement analyzed here as well as over the FIR in space-time requirement often posed in the discrete-time SLS results of e.g. \cite{wang2019system}. Further study of such alternate closed-loop constraints is the subject of future work.

 \subsection*{TF- and CLTF-Structure in QI Setting}
 Although CLTF-structure and TF-structure are comparable in the sense that they both bound the structured-realizable controller design problem (Thm.~\ref{thm:TF_CLTF_implySS}, Fig.~\ref{fig:networked_and_structured_realizability}), their performance is generally incomparable as emphasized in the following proposition, whose proof is in Appendix~\ref{app:CLTF_TF}.
 
 \begin{prop} \label{prop:CLTFvsTF}
		The sets of CLTF-structured and TF-structured controllers generally not comparable in the following sense: their intersection is non-empty and neither is a subset of the other. 
	\end{prop}
 
\begin{figure}[h!]
	\setlength\belowcaptionskip{-1.1\baselineskip}
	\centering 
			\includegraphics[width=0.42\textwidth]{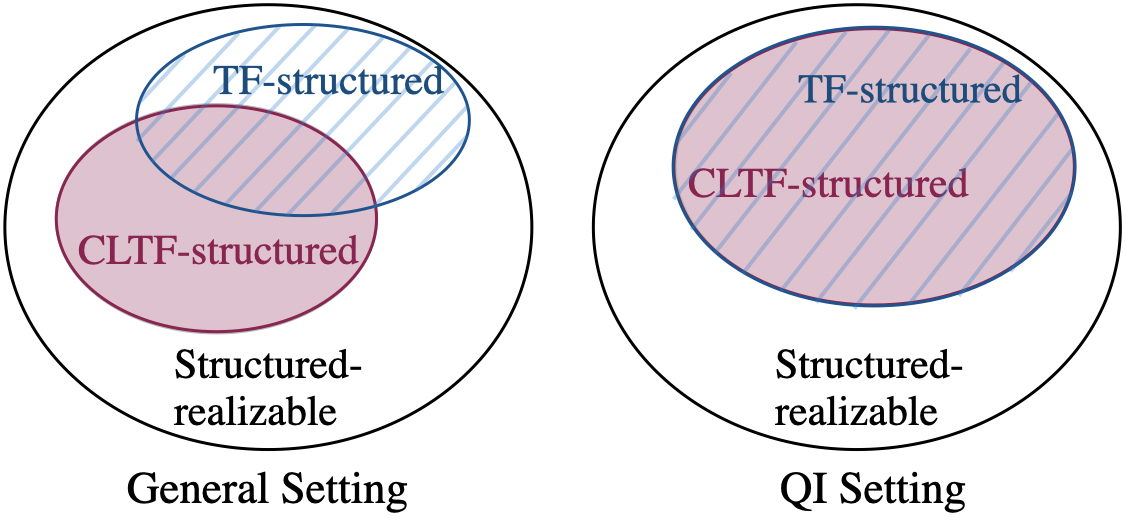} 
			
	\caption{\footnotesize Set relation between CLTF-structured, TF-structured, and structured-realizable controllers in (left) general settings and (right) in quadratically-invariant (or funnel causal) settings. Both sets are subsets of the set of structured-realizable controllers. The sets of CLTF-structured and TF-structured controllers are generally incomparable: their intersection is non-empty and neither is a subset of the other (left). Under quadratic invariance/ funnel causality, however, these sets are equivalent (right).} 
	\label{TFvsCLTF.fig}
\end{figure}

Quadratic invariance and funnel causality are two notable exceptions in which TF-structure and CLTF-structure align. The consensus example with added relative feedback constraints presented the opposite extreme of zero intersection of these two sets. Understanding the intersection of these two sets in more general settings is the subject of future work. \color{black}
These relations, and non-relations, are depicted in Figure~\ref{TFvsCLTF.fig}.

\subsection*{The Role of Minimality} 

Classical controller design is framed around the design of an optimal controller transfer function. There are many possible realizations of this transfer function, and in a classical or centralized  setting the obvious choice is to implement the minimal realization (unique up to similarity transformation). In a distributed setting, the choice of realization becomes more nuanced. If an admissible structured realization exists, it is likely not minimal. Thus, in certain settings there is an advantage to allowing more states/ memory if this leads to a distributed structure. 

We emphasize that there are many important open research directions in this setting. Examples include determining the state dimension needed to obtain a realization that is structured and admissible and understanding a potential tradeoff between local state dimension and the distributed structure. 
Answers to these and other questions are not obvious. A solution to the special two subsystem setting with lower-triangular structure was derived in \cite{lessard2015optimal}. It is unclear how such results generalize to more subsystems or alternate structures. Preliminary results for a more general number of subsystems and structures has largely focused on the stable setting \cite{vamsi2011optimal, vamsi2015optimal}. Realizations provided here are non-minimal and the possibility of reducing the dimension of these realizations or ensuring that they are admissible in the unstable setting remain as open questions. 

\section{Conclusion} 

In the context of distributed controller design, we formalized and compared three notions of controller locality and rigorously characterized relative feedback as a design constraint. We demonstrated that the sparse closed-loop design is a convex relaxation of the structured controller realization design problem, but the performance gap between these problems may be infinite (at least when relative feedback requirements are imposed). These results formalized and extended various works presented in the System Level Synthesis literature. 

An important open question is whether less restrictive convex constraints can be placed on the closed-loop to still ensure structured-realizability of the resulting controller. A more general open line of research is the characterization of admissible structured-realizable systems and an analysis of possible trade-offs between state dimension and the sparsity of the realization matrices in
 the distributed setting. The results in this paper motivate further study of these issues.

\color{black}

\bibliographystyle{ieeetr}
\bibliography{complete_bibliography}{}

\appendix

\subsection{Proof of Theorem~\ref{thm:TF_CLTF_implySS} (a) } \label{app:TFimpliesSS}

Let $G$ be TF-structured w.r.t. $\mathcal{A}$. We construct a $\cA$-structured realization of $G$ as follows. 
For all  $G_{ij} \not = 0$ , define $A_{ij}, B_{ij}, C_{ij}, D_{ij}$ from any realization $G_{ij} = C_{ij} (sI-A_{ij})^{-1}B_{ij}+D_{ij}$, 
and for $G_{ij} = 0$, define $A_{ij}, B_{ij},$ and $C_{ij}$ to be empty matrices, and $D_{ij} = 0$. 
Realize each row $G_i$ of $G$ as 
\begin{align*}
	G_i 
	&=  \left[ \begin{array}{c|c} \overline{A}_i & \overline{B}_i \\ \hline 
				\overline{C}_i & \overline{D}_i	\rule{0em}{1em} \end{array} \right]			{\small:=
	 \arraycolsep=3pt\def\arraystretch{1}\left[
	\begin{array}{c|c} 
		\begin{matrix} A_{i1} & & \\ & \ddots & \\ & & A_{i,\ssN} \end{matrix} & 
		\begin{matrix} B_{i1} & & \\ & \ddots & \\ & & B_{i,\ssN} \end{matrix}		\\ \hline
		\begin{matrix} C_{1i} & \cdots & C_{1,\ssN} \end{matrix}  	& 
			\begin{matrix} D_{1i} & \cdots & D_{1,\ssN} \end{matrix}  
	\end{array} 
	\right]	.}
\end{align*} 

The entire system $G$ can then be realized as 
\begin{multline*}
	\hspace{-.5em}
	G = 
	\left[ \begin{array}{c|c} A & B \\ \hline C & D \end{array} \right]		:=		
	\arraycolsep=3pt\def\arraystretch{1}
            	\left[
            	\begin{array}{c|c} 
            		\begin{matrix} \overline{A}_{1} & & \\ & \ddots & \\ & & \overline{A}_{\ssN} \end{matrix} & 
            		\begin{matrix} \overline{B}_{1}  \\  \vdots  \\  \overline{B}_{\ssN} \end{matrix}		\\ \hline
            		\begin{matrix} \overline{C}_{1} & \cdots & \overline{C}_\ssN \end{matrix}  	& 
            		\begin{matrix} \overline{D}_{1} & \cdots & \overline{D}_{\ssN} \end{matrix}  
            	\end{array} 
            	\right]	
				=																				\\
	{\footnotesize	\arraycolsep=3pt\def\arraystretch{1}
	\left[
	\begin{array}{ccc:c:ccc|c:c:c} 
		 {A}_{1,1} & 		&  				& 		& 				& 		&						& B_{1,1} 		& 		& 				\\
		 		& \ddots 	& 				&		& 				& 	 	&						& 				& \ddots 	& 				\\
		 		& 		& {A}_{1,\ssN} 	& 		& 				&		&						& 				&		& B_{1,\ssN}	\\ \hdashline
				&		&				&\ddots 	& 				&		&						&				& \vdots 	&				\\ \hdashline
		 	 	& 		&  				& 		& A_{\ssN,1}		& 		&						& B_{\ssN,1} 		& 		& 				\\
		 		& 	 	& 				&		& 				& \ddots 	&						& 				& \ddots 	& 				\\
		 		& 		& 			 	& 		& 				&		&A_{\ssN,\ssN}			& 				&		& B_{\ssN,\ssN}	\\ \hline						
		{C}_{1,1} & \cdots	& C_{1,\ssN}		&		&				&		&						& D_{1,1} 		& \cdots	& D_{1,\ssN}	\\ \hdashline
		 		&		&				& \ddots	&				&		&						& \vdots			&		& \vdots			\\ \hdashline				  
				& 		& 				&		&  C_{\ssN,1}	& \cdots	& C_{\ssN\sm 1,\ssN}			& D_{\ssN,1}		& \cdots	& D_{\ssN,\ssN} 	
	\end{array} 
	\right]	},
\end{multline*}
where the dashed lines represent the partitioning of inputs, outputs and states according to site index. It is clear that
$A$ and $C$ are block diagonal, and therefore $A,C\in\cS(\cA)$ trivially. 
 The matrices $B$ and $D$ have a block structure such that the $ij$'th block is zero 
if $\cA_{ij}=0$, i.e. $B,D\in\cS(\cA)$. Thus the realization $(A,B,C,D)$ is $\cA$-structured. Note that if $D(K) = D_K$ is block diagonal, then the above is a network-realization. 

A similar construction can alternatively yield a block-diagonal $B$ matrix if 
we begin by realizing each column of $G$ rather than each row.

\subsection{Proof of Theorem \ref{thm:TF_CLTF_implySS} (b)} 
\label{app:SLSimpliesDist}

We begin by stating and proving the following lemma, which demonstrates that the implementations of Figures~\ref{fig:sls_controller},~\ref{fig:sys_and_outputFB} do 
not introduce any unbounded internal signals when we restrict to  ${\rm Re}(\alpha) >0$. 

	\begin{lem} \label{lem:implementation}
	Assume the parameter $\alpha$ in block diagrams in Figures~\ref{fig:sls_controller},~\ref{fig:sys_and_outputFB} satisfies ${\rm Re}(\alpha) > 0$. Then, the following hold:
	\begin{enumerate} [label=(\alph*)]
		\item {\em (State Feedback)} If $\Phix, \Phiu$ are each stable, then all internal signals $\{ x, u, v \}$ in the block diagram of Figure~\ref{fig:sls_controller} are bounded whenever all disturbances signals $\{ d, n, w\} $ are bounded. 
		\item {\em Output Feedback} If $\Phi_{xx}, \Phi_{xy}, \Phi_{ux}, \Phi_{uy}$ are each stable, then all internal signals $\{x, u, y, v\}$ in the block diagram in Figure~\ref{fig:sys_and_outputFB} are bounded whenever all disturbance signals $\{w,d,n,m\}$ are bounded.
			\end{enumerate}
	\end{lem}
\begin{IEEEproof} To prove this result, we compute the transfer function from each exogenous disturbance to each internal signal. The results are summarized in the following two tables. 

\centerline{\it Closed-loop Transfer Functions - State Feedback:}
\begin{center}
\begin{tabular}{ c| l | l | l} 
~ &$w$&$n$ & $d$ \\ 
\hline
 $x$&$\Phix$ &  $\Phix(sI-A) - I$ & $\Phix B_2$  \\  
$ u$& $\Phiu$ & $\Phiu (sI-A)$& $I + \Phiu B_2$ \\
$v$ &$ \frac{1}{s+\alpha} $& $\frac{1}{s+\alpha}(sI-A)$ & $\frac{1}{s+\alpha}B_2$
\end{tabular}
\end{center}

~\\

\centerline{\it Closed-loop Transfer Functions - Output Feedback:}
\begin{center}
{\small
\begin{tabular}{ r | l | l | l | l} 
~ &$w$&$n$ & $d$ & $m$ \\ 
\hline
 $x$&$\Phix$ &  $\Phi_{xy}$ & $\Phi_{xx} B_2$ &$  \Phi_{xy} C_2$ \\ 
$ u$& $\Phiu$ & $\Phi_{uy}$& $ I + \Phi_{ux} B_2 $ & $\Phi_{uy} C_2 $\\
$v$ &$ \frac{1}{s + \alpha} C \Phi_{xx}$& $\frac{-1}{s+\alpha}B \Phi_{uy}$ & $\frac{-1}{s+\alpha} B_2 \Phi_{ux} B_2 $ & $ H(s)$\\
$y$ & $C \Phi_{xx}$ & $I + C \Phi_{xy}$ & $C_2 \Phi_{xx} B_2$ & $C_2 \Phi_{xy} C_2 $
\end{tabular}
}
\end{center}
where $H(s) := \frac{1}{s+\alpha}(sI-A)\left(I - \Phi_{xy} C_2\right)$. 
Each of these entries is stable under the assumption that $\Phi_{xx}, \Phi_{ux}, \Phi_{xy}, \Phi_{uy}$ are each stable.\end{IEEEproof}

We leverage these block diagrams to complete the proof.

{\em (1) State Feedback:}
Under the CLTF-structured assumption, $\Phix, \Phiu$ are TF-structured and strictly proper so that by the proof of Theorem ~\ref{thm:TF_CLTF_implySS} (a), there exist realizations of the form 
	\be \label{eq:Phi_real}
		\Phix = \lba{c|c} A_x & B_x \\ \hline C_x & 0 \ear, ~~ \Phiu = \lba{c|c} A_u & B_u \\ \hline C_u & 0 \ear,
	\ee 
with $A_x, A_u, C_x, C_u$ block diagonal and $B_x, B_u \in \mathcal{S}(\cA)$. We use this realization of $\Phix$ to construct a realization of $I - (s+\alpha) \Phix$:
	\be \ba
		I - (s+\alpha) \Phix &= \lba{c|c} A_x & B_x \\ \hline -C_x (A_x +\alpha I) & I - C_x B_x\ear\\
		& \overset{(1)}{=}  \lba{c|c} A_x & B_x \\ \hline -C_x (A_x +\alpha I) &0\ear
	\ea \ee
where equality (1) follows from the affine relation \eqref{CL_Constraints_state.eq} which shows that $I - (s+\alpha) \Phi_x = -(A+\alpha I) \Phix  - B \Phiu$ is strictly proper. Then the 
 feedback loop $\Psi_{xx}$ (see Figure~\ref{fig:sls_controller}) can be realized as
	\be\label{eq:psi_real}
		\Psi_{xx} = \lba{c|c} A_x - B_x C_x(A_x +\alpha I) & B_x \\ \hline -C_x(A_x +\alpha I) & I \ear =: \lba{c|c} A_{\psi} & B_{\psi} \\ \hline C_{\psi} & I \ear.
	\ee
Note that $A_{\psi}, B_{\psi} \in \mathcal{S}(\cA)$ and $C_{\psi}$ is block diagonal, so that \eqref{eq:psi_real} is a networked-realization. 
The controller $K$ is given by the cascade interconnection of $\Psi_{xx}$ with 
	\be
		(s+\alpha) \Phiu = \lba{c|c}  A_u & B_u \\ \hline A_u(C_u + \alpha I) & C_u B_u\ear,
	\ee
and can thus be realized as
	\be \label{eq:K_real1}
		K = \lba{cc|c} A_{\psi} & 0 & { B_{\psi}} \\
		B_u C _{\psi} & {A_u}  & B_u \\
		\hline 
		C_u B_u C_{\psi} &  A_u(C_u + \alpha I ) & C_u B_u
		 \ear.
	\ee
Each of the block components of the state matrices of realization~\eqref{eq:K_real1} are elements of $\mathcal{S}(\cA)$. Thus, after appropriate
re-ordering of the states and block partitioning of the matrices we recover a structured realization of $K$ which we denote by
	\be \label{eq:K_real}
		K = \lba{c|c}A_K & B_K \\ \hline C_K & D_K \ear.
	\ee

\begin{figure}[h!]
\begin{center}
\includegraphics[width=0.45\textwidth]{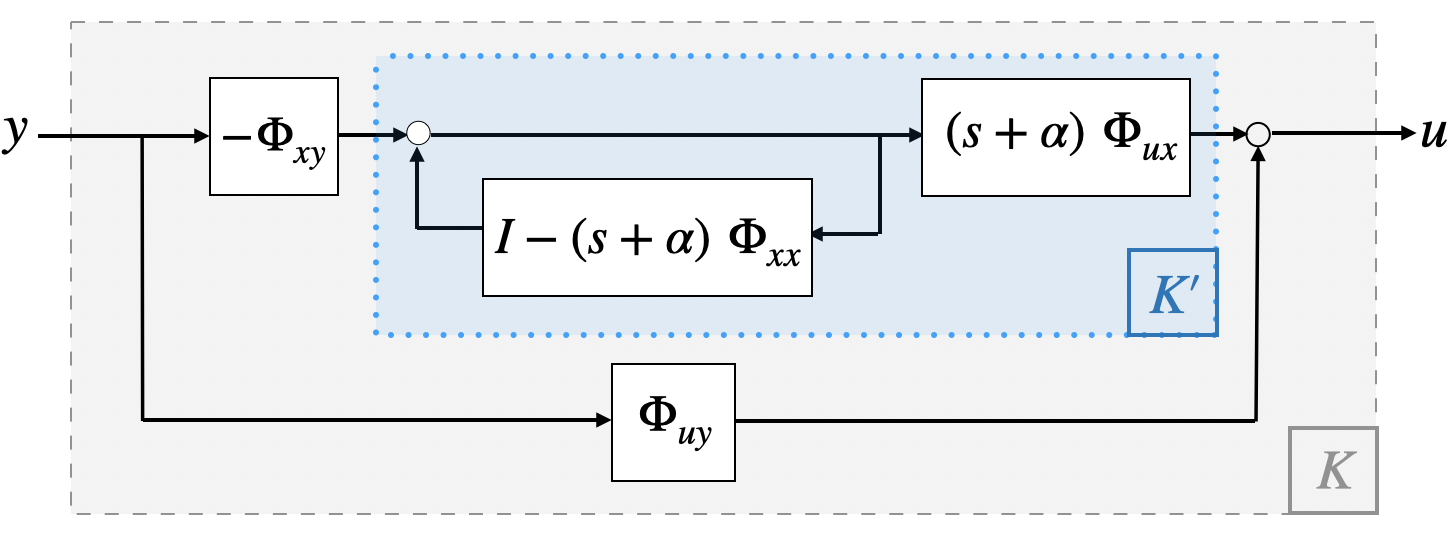} 
\end{center}
\caption{\footnotesize Block diagram implementation of the output feedback controller $K$ resulting in closed-loop mappings $\Phi_{xx}, \Phi_{xy}, \Phi_{ux}, \Phi_{uy}$. The scalar parameter $\alpha$ is chosen with ${\rm Re}(\alpha) > 0$ to insure internal stability of the interconnection of $K$ with plant $P$. }
\label{fig:output_fb_proof}
\end{figure}

{\em (2) Output Feedback:} In this case the controller is implemented as shown in Figure~\ref{fig:output_fb_proof}. Again $\Phi_{xx}$ and $\Phi_{ux}$ have realizations \eqref{eq:Phi_real}, and we 
recognize that the inner feedback loop, $K'$, (see Figure~\ref{fig:output_fb_proof}) will have the same structural properties as the controller $K$ in the state feedback setting, i.e. will have a realization of the form \eqref{eq:K_real}, which we denote by 
$K' = \lba{c|c} A_{K'} & B_{K'} \\ \hline C_{K'} & D_{K'} \ear$.

Since $\Phi_{xy}, \Phi_{uy}$ are TF-structured and $\Phi_{xy}$ is strictly proper, Theorem~\ref{thm:TF_CLTF_implySS} (a) implies they will have realizations of the form
	\be
		-\Phi_{xy} = \lba{c|c} A_{xy} & B_{xy} \\ \hline C_{xy} & 0 \ear, ~~ \Phi_{uy} = \lba{c|c} A_{uy} & B_{uy} \\ \hline C_{uy} & D_{uy} \ear,
	\ee
with $A_{xy}, A_{uy}, C_{xy}, C_{uy}$ block diagonal and $B_{xy}, B_{uy}, D_{uy} \in \mathcal{S}(\cA)$. One realization of the cascade interconnection of $\Phi_{xy}$ and $K'$ is then given by 
	\be \label{eq:cascade_real}
		K' \Phi_{xy} = \lba{cc|c}A_{xy} & 0 & B_{xy} \\ B_{K'} C_{xy} & A_{k'} & 0 \\ \hline D_{K'} C_{xy} & C_{K'} & 0 \ear
	\ee
	Each block entry of each matrix in the realization \eqref{eq:cascade_real} is an element of $ \in \mathcal{S}(\cA)$, so that after appropriate
re-ordering of the states and block partitioning of the matrices we recover a structured realization of $K' \Phi_{xy}$ which we denote by 
	\be
		K' \Phi_{xy}  = \lba{c|c}A_c & B_c \\ \hline C_c & 0 \ear
	\ee
Forming $K$ as the feedback interconnection of $K' \Phi_{xy}$ and $\Phi_{uy}$ provides the realization 
	\be \label{eq:K_real_output}
		K = \lba{cc|c} A_c & 0 & B_c \\0 & A_{uy} & B_{uy} \\ \hline C_c & C_{xy} & D_{uy}  \ear 
	\ee
Again re-ordering of states and and block partitioning of realization matrices leads to a structured realization which we denote as
	\be\label{eq:K_real_output2}
		K = \lba{c|c} A_K & B_K \\ \hline C_K & D_K\ear.
	\ee

We next build off these results to prove the network realizability statements. 

{\it Proof of statement (b - i):} 
If $K$ is CLTF-structured, then there exists a structured realization of $K$ of the form \eqref{eq:K_real1}. From this realization, we see that the direct feedthrough term of $K$ is equal to the direct feedthrough term of $s\Phiu$:
	\be
		D(K) = C_u B_u = D(s\Phiu). 
	\ee
Thus, a necessary condition for network-realizability of $K$ is that $D(s\Phiu)$ is block diagonal. Conversely, if $D(s\Phiu)$ is block diagonal, then \eqref{eq:K_real1} is a networked realization of $K$. 

{\it Proof of statement (b - ii):} 
First suppose $K$ is CLTF-structured. Then there exists a structured realization of $K$ of the form \eqref{eq:K_real_output}. From this realization, we see that the direct feedthrough term of $K$ is equal to that of $\Phi_{uy}$:
	\be
		D(K) = D(\Phi_{uy}).
	\ee
Thus, a necessary condition for network-realizability of $K$ is that $D(\Phi_{uy})$ is block diagonal. Next suppose $D(\Phi_{uy})$ and $D(s\Phi_{ux})$ are both block diagonal. To show that under this assumption \eqref{eq:K_real_output2}  is a networked realization of $K$, we show that $C_K$ of \eqref{eq:K_real_output2} is block diagonal when $D(\Phi_{uy})$ and $D(s\Phiu)$ are both block diagonal. 
$C_K$ is block diagonal if $D_{uy}, C_{xy}$ and the block entries of $C_c$, $D_{K'} C_{xy}$ and $C_{K'}$, are each block diagonal. CLTF-structure implies block diagonality of $C_{xy}$ and $C_{K'}$. $D_{K'} = D(s \Phi_{ux})$.

\subsection{Proof of Proposition~\ref{prop:notTF_CLTF}} \label{app:notTF_CLTF}
\begin{IEEEproof}
(a) Define $K = \left[ \begin{array}{c|c} A & I \\ \hline I & 0 \end{array} \right],$ with
	%
	 $A:= \left[ \begin{array}{ccc} \sm2 & 1 & 0 \\ 1 & \sm2 & 1 \\ 0 & 1 & \sm2 \end{array} \right].$ This is an admissible $\cA$-structured (also $\cA$-networked) realization for graph $\cA$ defined by 			
	 \be \label{eq:A}
	\cA := \left[ \begin{array}{ccc} 1 & 1 & 0 \\ 1 & 1 & 1\\ 0 & 1 & 1 \end{array} \right],
	\ee and a direct calculation shows that $K (s)= I(sI-A)^{-1}I$
is not TF-structured w.r.t. $\cA$. \\
(b)  Consider the admissible network-realizable controller $K_a$ defined in \eqref{eq:Ka} in feedback with plant \eqref{eq:plantState}. The corresponding closed-loop $\Phix = (sI-0 - I \cdot K_a(s))^{-1}$ is a full transfer function matrix so that $K_a$ is not CLTF-structured.
\end{IEEEproof}

\subsection{Proof of Theorem~\ref{lem:relativeK}} 
For simplicity of notation in this proof, we will often drop the argument $s$ from transfer functions.  
The first direction of statement (a) is immediate.  If 
$K$ is relative, meaning that it can be written in the form~\req{relative_def}, then 
\[
	 K_m \mbo ~=~ 
	 \sum_{1\leq i < j \leq N}  \cK^{(m)}_{ij} ~  (I-I) ~=~ 0.
\]

The other direction is to show that if $K_m\mbo=0$, then we can rewrite $K_m ~\!y$ in the form 
\be
		u_m ~=~ K_m~y ~=~ \sum_{1\leq i < j \leq N}  \cK^{(m)}_{ij} ~  (y_i - y_j). 
   \label{sum_cK.eq}
\ee
For simplicity of notation, we will drop the superscript $m$ in the remainder of the proof. 
In addition, we will assume without loss of generality  that $u$ is scalar valued, and that each $\cK_{ij}$ is SISO, so that each signal $y_i$ 
is scalar-valued. If $Ky$ can be written in the above form for each of the scalar subcomponent $y_i$, then concatenating  
these representations as columns would give the representation for vector signals $\{y_i\}$. 

The form~\req{sum_cK} involves $(N^2-N)/2$ transfer functions $\{ \cK_{ij} \}$. 
 Form the $N\times N$ skew-symmetric (not skew-Hermitian)
 transfer function  matrix 
\be
	\cK :=  
			 \bbm 0 &  \sm\cK_{12} & \cdots 		& \sm\cK_{1\ssN}	\\
				\cK_{12} & 0 & 			& 		\\ 
				\vdots 		& 	& 	\ddots	& 		\\ 
				 \cK_{1\ssN} & 	& 			& 0
			\ebm .
  \label{cK_def.eq}
\ee
Then the relation~\req{sum_cK} can be written more  compactly as
\[
	Ky 
		= \mbo^\ssT   \cK y . 
\]	
Therefore the relation  between $K$ and $\cK$ (after transposing)  is 
\be 
	 \bbm 0 &  &  \cK_{ij}	\\
		& \ddots  & 		\\ 
	\sm \cK_{ij} & 	& 0	\ebm
	\bbm 1 \\ \vdots \\ 1 \ebm 
	 =
	 \bbm K_1 \\ \vdots \\  K_\ssN \ebm .
  \label{cK_K.eq}
\ee
\eqref{cK_K.eq} is
a highly underdetermined system of linear equations so that if one solution $\cK$ exists for a a given $K$, then there are an infinite number of other solutions. To prove  
Part (b) of the theorem statement, we characterize when there exists solutions with a particular sparsity structure, i.e. where the pairings $(i,j)$ for which $\cK_{ij}$ are nonzero in representation \req{sum_cK} are 
selected as the edges of a pre-specified graph:

	 Define the following sets of complex matrices 
	 \begin{align*} 
	 	\skw &:= \lcb M\in \C^{\ssN\times\ssN}; ~M^\ssT \!= \sm M \rcb 	,~ \mbox{\footnotesize (skew-symmetric matrices)}	\\
	 	\sA 	&:= \lcb M\in \skw; ~~\spM{M} = \cA  \rcb 				, ~\mbox{\footnotesize (skew-symmetric w/ sparsity $\cA$)}, 
	 \end{align*}
	 where $\spM{M}$ stands for the matrix of the sparsity pattern of $M$. 
	 Note that these sets are vector spaces. 
	 Now consider the matrix operator 
	 \[
	 	\cL_{\sA}: \sA \rightarrow \C^\ssN, 
		\hspace{2em} 
		\cL_{\sA}\lb M \rb := M\mbo . 
	 \]
	 The solvability of 
	 \be
	 	\cK^\ssT(s) ~\mbo ~=~ K^\ssT(s) ,  
	  \label{solvability.eq}
	 \ee
	with a solution of same sparsity as $\cA$
	is equivalent to the solvability of 
	 \[
	 	\cL_{\sA}\lb \cK(s) \rb ~=~ K^\ssT(s) , 
	 \]
	 which in turn is equivalent to the statement 
	 \be
	 	K^\ssT(s)  ~\in~ \Ims{ \cL_{\sA}} , 
	  \label{K_in.eq}
	 \ee
	 where $\Ims{\cdot}$ denotes the image space of the operator. All of the above statements are to be interpreted as required
	 to hold  for each $s\in\C$ except isolated points. 
	 
	 Thus we have converted the solvability question to one about the range space of a matrix operator. From the fundamental theorem of linear algebra 
	 \[
	 	\Ims{ \cL_{\sA} } ~=~ \Ims{ \cL_{\sA}  \cL_{\sA}^\dagger},
	 \]
	 where $\cL_{\sA}^\dagger:\C^\ssN\rightarrow \C^{\ssN\times\ssN}$ is the adjoint. 
	 It is much easier to characterize $\cL_{\sA}  \cL_{\sA}^\dagger$ in terms of the graph connectivity
	 as stated in the  following 
	 lemma whose proof is stated at the end of this section.
	 \begin{lem} 													\label{LLdagger.lem}
	 	The composition of $\cL_{\sA}$ with its adjoint is given by 
		\[
			 \cL_{\sA}  \cL_{\sA}^\dagger ~=~\frac{1}{2} ~ L 
		\]
		where $L$ is the Laplacian of the graph $\cA$. 
	 \end{lem} 
	 
	 For undirected networks, $L$ is a symmetric matrix, and thus its image and null spaces are mutually orthogonal. 
	 A standard result in algebraic graph theory states that a
	  graph is connected iff the null space of $L$ is just $\mbo$, i.e. it is connected iff 
	 \[
	 	\Nus{L}={\rm span}(\mbo)
		\hspace{2em} \Leftrightarrow \hspace{2em} 
		\mbo^\perp = \Ims{L}. 
	\]
	Note that the condition~\req{K_in} is required not for any $K(s)$, but only those that are such that $K(s)\mbo=0$, i.e. 
	$K(s)\in \mbo^\perp$, which is exactly $\Ims{L}$. We therefore conclude that 
	\[
		K \mbo=0
		~~\Rightarrow~~
		K^\ssT \in \Ims{L} = \Ims{\cL_{\cA} \cL_{\cA}^\dagger} = \Ims{\cL_{\cA}}, 
	\]	
	and \req{solvability} is solvable with a $\cK$ that has the same sparsity structure as the graph $\cA$. This proves Theorem~\ref{lem:relativeK}. \hfill $\blacksquare$

\subsubsection*{Proof of Lemma~\ref{LLdagger.lem}} \label{relative.appendix}

The first step is to compute the adjoint $\cL_{\cS_\cA}^\dagger$, and then compute the composition 
$\cL_{\cS_\cA} \cL_{\cS_\cA}^\dagger$. 
To compute the adjoint, it is easier to work with the following operator 
	 \[
	 	\cL: \C^{\ssN\times\ssN}  \rightarrow \C^\ssN, 
		\hspace{2em} 
		\cL\lb M \rb := M\mbo , 
	 \]
and note that $\cL_{\cS_\cA}= \left. \cL\right|_{\cS_\cA}$, i.e. the restriction of $\cL$ to $\cS_\cA$. It then follows that 
\[
	\cL_{\cS_\cA}^\dagger = \Pi_{\cS_\cA} \cL^\dagger  =  \Pi_{\cS_\cA} \Pi_{\cS}  \cL^\dagger, 
\]
where we have written the projection as the composition of two projections that are each easier to compute. 
In summary, 
\[
	\cS_\cA
	\stackrel{\Pi_{\cS_\cA}}{\longleftarrow} 
	\cS
	\stackrel{\Pi_{\cS}}{\longleftarrow} 
	\C^{\ssN\times\ssN} 
	\stackrel{\cL^\dagger}{\longleftarrow} 
	\C^\ssN 
\]
If $M$ is a skew-symmetric matrix, then 
\[
	\Pi_{\cS_\cA}:\cS \rightarrow \cS_\cA, 
	\hspace{2em} 
	\Pi_{\cS_\cA} \lb M \rb = \cA \circ M , 
\]
where $\circ$ is the Hadamard (element-by-element) product of two matrices. 
Now if $M$ is any complex matrix, then
\[
	 \Pi_{\cS}: \C^{\ssN\times\ssN} \rightarrow \cS, 
	 \hspace{2em} 
	 \Pi_{\cS} \lb M \rb = \frac{1}{2} \lb M - M^\ssT \rb . 
\]
Finally, given any complex vector $v$  
\[
	\cL^\dagger: \C^\ssN \rightarrow \C^{\ssN\times\ssN} , 
	\hspace{2em} 
	\cL^\dagger\lb v \rb = v \mbo^\ssT. 
\]
The last fact follows from the requirement $\forall M\in \C^{\ssN\times\ssN}$ 
\[
	{\rm tr}\lb ( \cL^\dagger v )^* M \rb = 
	\inprod{\cL^\dagger v}{M} \equiv \inprod{v}{\cL(M)} 
	= {\rm tr}\lb v^* M\mbo \rb .
\]
Putting it all together, we conclude that 
\[
	\cL_{\cS_\cA}^\dagger v = \Pi_{\cS_\cA}\Pi_{\cS} \cL^\dagger v 
	= \frac{1}{2}~ \cA \circ  \lb v\mbo^\ssT - \mbo v^\ssT \rb 
\]

Finally, we compute the composition $\cL_{\cS_\cA} \cL_{\cS_\cA}^\dagger$. First note that 
if $\cA$ is the adjacency matrix of an undirected graph, the following is a useful 
characterization of the  Laplacian 
\[
	L = D - \cA = \diag{\cA \mbo} - \cA , 
\]	
where $D$ is the diagonal matrix of node degrees, which is $\diag{\cA\mbo}$, where $\diag{w}$ 
makes a diagonal matrix from the entries of the vector $w$. Now compute 
\begin{align*} 
	2 ~\cL_{\cS_\cA} \cL_{\cS_\cA}^\dagger v 
	&=  
	\big[ \cA \circ \lb v\mbo^\ssT -\mbo v^\ssT \rb \big] \mbo		\\
	&=  
	\big[ \cA \circ \lb v\mbo^\ssT\rb  - \cA \circ \lb \mbo v^\ssT \rb \big] \mbo	\\
	&\stackrel{1}{=}  
	\big[ \diag{v}  \cA  ~\diag{\mbo}    - \diag{\mbo} \cA~ \diag{v}   \big] \mbo	\\
	&=  
	\big[ \diag{v}  \cA  ~I   - I \cA~ \diag{v}   \big] \mbo						\\
	&=  
	\diag{v}  \cA  ~\mbo  -  \cA~ \diag{v}    \mbo						\\
	&\stackrel{2}{=}  
	\diag{\cA \mbo}  v  -  \cA~ v	~=~ Dv - \cA v ~=~ L v, 
\end{align*} 
where $\stackrel{1}{=}$ follows from the fact that the Hadamard product with a rank one matrix 
$\cA\circ (uw^\ssT) = \diag{u} \cA ~\diag{w}$, and $\stackrel{2}{=}$ follows from 
$\diag{v} w = \diag{w} v$ for any two vectors $v$, $w$.

\subsection{Proof of Proposition~\ref{prop:relative_realization}} \label{app:relative_realization}
If  the transfer function $K$ is relative, then each entry of the impulse response of $K$ must be relative, i.e. $D \mathbb{1} = 0$ and $C e^{At} B \mathbb{1} = 0$ for all $t$. Then 
	\be 
		W_o(t) (B \mathbb{1} ) = \int_0^t e^{A^T \tau} C^T C e^{A\tau}B \mathbb{1} d \tau = 0. 
	\ee
If $(A,C)$ is observable then the observability Gramian $W_o$ is full rank, so that $W_o(t) (B \mathbb{1} ) = 0$ implies $B \mathbb{1} = 0$. \hfill $\blacksquare$

\subsection{Proof of Theorem \ref{lem:relative}} \label{relativeSLS.appendix}

Given a controller $u = Kx$, the resulting closed-loop transfer function $\Phiu$ for system \eqref{eq:PlantDynamics} is given by
 	$$
		\Phiu(s) = K(s) \Phix(s) =  K(s) (sI-A - BK(s))^{-1}.
	$$
(a)  First assume $K$ is relative, i.e. $K(s) \mathbb{1} = 0$. Then 
	$$	
		\Phix^{-1}(s) \one = (sI-A - BK(s)) \mathbb{1} = s \mathbb{1}
	$$
	\begin{equation} \begin{aligned}
		\Rightarrow ~\Phiu (s)\mathbb{1} ~&= K(s) \Phix(s) \mathbb{1}= \frac{1}{s} B \cdot K (s) \mathbb{1} =0.
	\end{aligned} \end{equation}
Conversely, if $\Phiu$ is relative, i.e. $\Phiu(s) \mathbb{1} = 0,$ then
\begin{equation*}\begin{aligned}
		\mathbb{1} &= (sI-A - BK(s)) (sI-A - BK(s))^{-1} \mathbb{1}\\
		&= (sI-A)(sI-A-BK(s))^{-1}\mathbb{1} - B \Phiu(s) \mathbb{1}\\
		& =  (sI-A)\Phix(s)\mathbb{1}.
	\end{aligned} \end{equation*}
	Thus, using the fact that $A$ is relative,
		\begin{equation*} \begin{aligned}
		0 & = (sI-A)^{-1}  \mathbb{1}- \Phix\mathbb{1}
		= \frac{1}{s} \mathbb{1} - \Phix \mathbb{1}.
	\end{aligned} \end{equation*}
Rearranging, we have that
	\begin{equation*} \begin{aligned}
		&s \mathbb{1} = \Phix^{-1}(s) \mathbb{1} = s \mathbb{1} - B K(s) \mathbb{1}~~\Rightarrow~~ B K(s) \mathbb{1} = 0,
	\end{aligned}\end{equation*}
which implies (since $B$ is full rank) that $K\mathbb{1} = 0$. \\
(b)  First assume $K$ satisfies \eqref{eq:partitionedKconstraint}. For $A, B_2$ of the form \eqref{eq:ABpartitioned}, 
	\be\label{eq:Phixpartitioned}
		\Phix^{-1}(s) = \lba{cc} sI - A_1 & -I \\ -A_2-\overline{B} K_1 & sI - A_3 - \overline{B} K_2 \ear
	\ee
and it is straightforward to compute 
	\be \label{eq:PhixOnePartitioned}
		\Phix^{-1} \lba{c} \one \\ 0 \ear = s \lba{c} \one \\ 0 \ear, ~~ \Phix^{-1} \lba{c} 0 \\ \one \ear = \lba{cc} - \one \\ s \one \ear.
	\ee
Left multiplying \eqref{eq:PhixOnePartitioned} by $\Phix$ gives
	$
		\Phix \lba{c} \one \\ 0 \ear = \lba{c} \frac{1}{s} \one \\ 0 \ear, ~\Phix \lba{c} \one \\ \one \ear = \lba{c}0 \\  \frac{1}{s}\one  \ear. 
	$
By linearity, 
	$
		\Phix \lba{c} 0 \\ \one \ear =\frac{\sm1}{s} \lba{c} \one \\  \one \ear. 
	$
Then, 
	$$
		\Phiu(s) \lba{c} \one \\ 0 \ear = \lba{cc} K_1(s) & K_2(s) \ear \Phix(s)  \lba{c} \one \\ 0 \ear = 0,
	$$
and similarly $\Phiu(s) \lba{c}0 \\ \one\ear = 0$.  The proof of the converse follows similarly to that of statement (a). \hfill $\blacksquare$
\color{black}

\subsection{Proof of Proposition~\ref{prop:Ka}} \label{app:propKa}
Using representation \eqref{eq:cl} for the closed-loops, we compute
	\be \ba
		&\| \F(P; K_a) - \F(P;K_s) \|  \\
		 &\le \underbrace{\|P_{12}\big( (I- P_{22} K_a)^{\sm1} - (I- P_{22} K_s)^{\sm1} \big) \|}_{\text{( \MakeUppercase{\romannumeral 1} )}} \cdot \underbrace{\| P_{21}\|}_{\text{( \MakeUppercase{\romannumeral 2} )}},	
		 \ea \ee
	where each $\|\cdot \|$ denotes the $\h_2$ norm. 
Since (\MakeUppercase{\romannumeral 2}) is bounded, it's sufficient to prove that (\MakeUppercase{\romannumeral 1}) $\rightarrow 0$ as $a \rightarrow - \infty$. (\MakeUppercase{\romannumeral 1}) can be written equivalently as 
	\be \ba
	&\left\| 
		 	\begin{array}{l}
			P_{12}  (I\sm K_sP_{22})^{\sm 1} \Big( \left( I\sm K_sP_{22} \right)K_a\\
			 ~~~~~~~~~~~~~~~~~~~~~~ -   K_s\left (I\sm P_{22}K_a\right)\Big ) (I\sm P_{22}K_a)^{\sm1}
			\end{array}
		\right\|\\
	& = \left\|
			P_{12} ~ (I\sm K_sP_{22})^{\sm 1} ~\Big(  K_a - K_s \Big)~  (I\sm P_{22}K_a)^{\sm1}
		\right\| \\
	& \le \underbrace{\left\|P_{12}  ~(I\sm K_sP_{22})^{\sm 1}  \right\|}_{< \infty} \cdot \underbrace{\left\| (K_a - K_s) \cdot (I \sm P_{22}K_a)^{\sm 1} \right\|}_{\text{( \MakeUppercase{\romannumeral 3} )}}
	\ea \ee
To see that (\MakeUppercase{\romannumeral 3}) $\rightarrow 0$ as $a \rightarrow - \infty$, note that the low pass filter $K_a$ is essentially a static gain equal to $K_s$ up until frequency $|a|$ and $(I \sm P_{22}K_a)^{\sm 1}$ decays at high frequency since it is strictly proper. 

\subsection{Completion of Proof of Theorem~\ref{thm:infeasibility} \label{app:proofCompletion}}
Assume the cost is finite. Then $\frac{1}{s} C ( I - \Phiu)$ must be stable. Then each entry of the transfer matrix $C ( I - \Phiu(s) )$ has a zero at $s = 0$. Equivalently, for each $i = 1, ..., n$, 
	\begin{equation} \label{eq:ci}
		C_i - C (\Phiu)_i(0) = 0_{n \times 1}
	\end{equation}
where $C_i$ is the $i^{\text{th}}$ column of $C$, and $(\Phiu)_i$ is the $i^{\text{th}}$ column of $\Phiu$. Define a mapping
	 $$
	 	(\Phiu)_i(0) \in \R^{n \times 1} \mapsto (\tilde{\Phi}_u)_i(0) \in \R^{(2b+1) \times 1}
	$$
which removes all constrained zero entries of $(\Phiu)_i$ due to the constraint that $\Phiu$ is TF-structured w.r.t. $\cA^{(b)}$. Similarly define a mapping 
	$$
		C \in \R^{n \times n} \mapsto \tilde{C}(i) \in \R^{n \times (2b+1)}
	$$
 by extracting the columns of $C$ which correspond to the constrained zero entries of $\Phiu$. Then \eqref{eq:ci} can be rewritten as 
	\begin{equation} \label{eq:ciTilde}
		\tilde{C}(i) (\tilde{\Phi}_u)_i(0) = C_i. 
	\end{equation}
One solution $\tilde{\Phi}^u_i(0)$ of \eqref{eq:ciTilde} is given by the unit basis vector $e_k$, where $k$ denotes the column of $\tilde{C}(i)$ that is equal to $C_i$. 
Since $C$ is circulant and of rank $r > (2b+1)$, the matrix $\tilde{C}(i)$ will have full column rank, so that this solution $(\tilde{\Phi}_u)_i(0) $ is unique. Thus, the solution $\Phiu(0)$ composed of all columns $(\Phiu)_i$ will be nonzero and will contain entries of only ones and zeros. Thus, it could not be that
$\Phiu \mathbb{1} = 0$.

\subsection{Proof of Proposition~\ref{prop:CLTFvsTF}} \label{app:CLTF_TF}
We first prove this result through counterexamples. \\
1) The controller 
{\small
	\begin{equation*} \begin{aligned}
		&K(s) = \frac{s}{(s+1)^2 (s+2)^2 - (s+1)^2 - (s+2)^2} ~\cdot \\
		& \left[ \begin{array}{ccc} (s+2)^2 & (s+1)(s+3)^2 & - (s+1)(s+3) \\ (s+1)(s+3)^2 & -(s+1)^2 - (s+3)^2 & (s+1)^2(s+3) \\ -(s+1)(s+3) & (s+1)^2(s+3) & -(s+1)^2 \end{array} \right]
	\end{aligned} \end{equation*}
}
is not TF-structured w.r.t. the graph $\cA$ defined in Eq. \eqref{eq:A} since $K_{31}, K_{13} \ne 0$. For plant dynamics $\dot{x} = u + w$ the closed-loop maps resulting from $K$ are given by 
	$${\small
		\Phix= \frac{1}{s} \left[ \begin{array}{ccc} 1 & \frac{1}{s+1}& 0 \\ \frac{1}{s+1} & 1 & \frac{1}{s+2} \\ 0 & \frac{1}{s+2} & 1 \end{array} \right],
		\Phiu=\left[ \begin{array}{ccc} 0 & \frac{1}{s+1} & 0 \\ \frac{1}{s+1} & 0 & \frac{1}{s+2} \\ 0 & \frac{1}{s+2} & 0  \end{array} \right]
		}
	$$
so that $K$ is CLTF-structured for $\cA$ defined in Eq. \eqref{eq:A}.\\
2) The static controller $K = \cA$ is clearly TF-structured w.r.t. $\cA$. Compute
	\begin{equation*} \ba
		 \Phi^x &= (sI \sm K)^{-1} 
		 = \frac{-1}{s^2 - 2s -1}\arraycolsep=3pt\def\arraystretch{1} \lba{ccc}
			\frac{-s^2 + 2s}{s-1} & 1 & \frac{1}{s-1}\\
			1 & \sm s+1 & 1\\
			\frac{1}{s-1} & 1 & \frac{-s^2 + 2s}{s-1}
		 \ear
	\ea \end{equation*}
to see this controller is not CLTF-structured for $P$ w.r.t. $\cA$.



\ifCLASSOPTIONcaptionsoff
  \newpage
\fi



%

%

\begin{IEEEbiography}[{\includegraphics[width=1in,height=1.25in,clip,keepaspectratio]{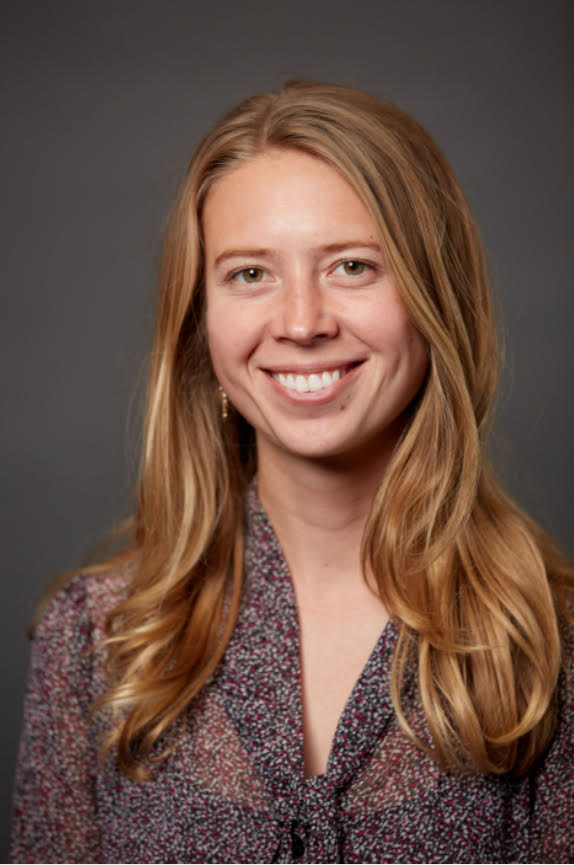}}]{Emily Jensen}
received the B.S. degree in engineering mathematics and statistics from the University of California, Berkeley, CA, USA, in 2015, and the M.S. and Ph.D. degrees in electrical and computer engineering from the University of California, Santa Barbara (UCSB), CA, USA, in 2019 and 2020, respectively. 
She was a Research Assistant with the Department of Computing and Mathematical Sciences, Caltech, Pasadena, CA, USA, until beginning her graduate studies in 2016. She is
currently a Postdoctoral Researcher with the Mechanical and Industrial Engineering Department, Northeastern University, Boston, MA, USA. Dr. Jensen was the recipient of the UC Regents’ Graduate Fellowship in 2016, and of the Zonta Amelia Earhart Fellowship in 2019.
\end{IEEEbiography}

\begin{IEEEbiography}[{\includegraphics[width=1in,height=1.25in,clip,keepaspectratio]{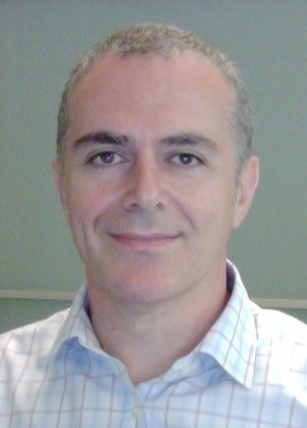}}]{Bassam Bamieh}
(F’08) received the B.Sc. degree in
electrical engineering and physics from Valparaiso
University, Valparaiso, IN, USA, in 1983, and the
M.Sc. and Ph.D. degrees in electrical and computer
engineering from Rice University, Houston, TX,
USA, in 1986 and 1992, respectively. From 1991
to 1998 he was an Assistant Professor with the
Department of Electrical and Computer Engineering,
and the Coordinated Science Laboratory, University
of Illinois at Urbana-Champaign, after which he
joined the University of California at Santa Barbara
(UCSB) where he is currently a Professor of Mechanical Engineering. His research interests include robust and optimal control, distributed and networked
control and dynamical systems, shear flow transition and turbulence, and the
use of feedback in thermoacoustic energy conversion devices. He is a past
recipient of the IEEE Control Systems Society G. S. Axelby Outstanding Paper
Award (twice), the AACC Hugo Schuck Best Paper Award, and the National
Science Foundation CAREER Award. He was elected as a Distinguished
Lecturer of the IEEE Control Systems Society (2005), Fellow of the IEEE
(2008), and a Fellow of the International Federation of Automatic Control
(IFAC).
\end{IEEEbiography}








\end{document}